\documentclass[]{aa}
\usepackage[dvips]{graphicx}
\usepackage{natbib}
\usepackage{txfonts}
\def\ebv{\mbox{$E_{B-V}$\,}}
\def\halpha{\mbox{H$\alpha$}}

\def\peryr{\mbox{$\>\rm yr^{-1}$}}

\def\subsun{\mbox{$_{\normalsize\odot}$}}
\def\lesssim{\mathrel{\hbox{\rlap{\hbox{\lower4pt\hbox{$\sim$}}}\hbox{$<$}}}}

\mail{lchristensen@aip.de}
\begin{document}

\title{UV star-formation rates of GRB host galaxies}

\author{L. Christensen\inst{1}
  \and J. Hjorth\inst{2}
  \and J. Gorosabel\inst{3,4,5} 
}
\institute{Astrophysikalisches Institut Potsdam, An der Sternwarte 16, 
14482 Potsdam, Germany
 \and Niels Bohr Institute, Astronomical Observatory, Juliane Maries Vej 30, 
DK--2100 Copenhagen, Denmark
 \and Danish Space Space Research Institute,
Juliane Maries Vej 30, DK--2100 Copenhagen~\O, Denmark
 \and  Space Telescope Science Institute, 3700, San Martin Drive, Baltimore, 
  MD 21218, USA
 \and Instituto de Astrof\'{\i}sica de Andaluc\'{\i}a, IAA-CSIC, Granada, Spain
}

\date{Received / Accepted}

\abstract{ We study a magnitude-limited sample of 10 gamma-ray burst (GRB)
  host galaxies with known spectroscopic redshifts ($0.43 < z < 2.04$).  From
  an analysis of the spectral energy distributions (SEDs), based on published
  broad-band optical and near-infrared photometry, we derive photometric
  redshifts, galaxy types, ages of the dominant stellar populations, internal
  extinctions, and ultraviolet (UV) star-formation rates (SFRs) of the host
  galaxies. The photometric redshifts are quite accurate despite the
  heterogeneous nature of the sample: The r.m.s. errors are $\sigma(z) = 0.21$
  and $\sigma(\Delta z/(1+z_{\rm spec})) = 0.16$ with no significant
  systematic offsets.  All the host galaxies have SEDs similar to young
  starburst galaxies with moderate to low extinction. A comparison of specific
  SFRs with those of high-redshift galaxies in the Hubble Deep Fields shows
  that GRB hosts are most likely similar to the field galaxies with the
  largest specific SFRs. On the other hand, GRB hosts are not significantly
  younger than starburst field galaxies at similar redshifts, but are found to
  be younger than a sample of all types of field galaxies.
   } \maketitle

\section{Introduction}
The association of Gamma-Ray Bursts (GRBs) with collapsing massive stars has
been inferred through many observations during the past 7 years, notably
through supernova bumps in the afterglow light curves or localisations of the
afterglow close to star forming regions
\citep{ctir99,bloom99,fruchter99,gal98,kul98b,holland99b,fynbo00}.
Additionally, \citet{bloom00} argued that GRB progenitors, given their small
positional offsets relative to their hosts, are likely massive stars.
Evidence that at least some long-duration GRBs are associated with supernovae
came with the observations of the afterglow of GRB 030329, which showed
distinct spectral supernova features \citep{stanek03,hjorth03}.

Underlying host galaxies have been found in all cases where GRBs were
localised with sub-arcsecond precision. Currently, the sample consists of
$\sim$35 such hosts with known redshifts in the range $0.1055<z<4.5$
\citep{prochaska04,andersen00}. The reported magnitudes of the hosts are
$20.4~<~R~<~30$ \citep{malesani04,jaunsen03}.

The life time of the massive stars believed to produce long-duration GRBs is
of the order of a few Myr. If the host galaxies are indeed forming such
massive stars this should be reflected in their spectral energy distributions
(SEDs) which, in the absence of reddening, should reveal a blue continuum.
Moreover, their star-formation rates (SFRs) should be large. The integrated
SFR of a galaxy can be found from measurements of \halpha \, or [\ion{O}{II}]
line fluxes, or from measuring the flux in the UV continuum at
1500--2800~{\AA} in the rest frame of the galaxy \citep{kennicutt98}.  The
faintness of GRB hosts presents a problem for spectroscopy as they require
long integration times on the largest telescopes. Ground-based photometry in
several filters presents an alternative possibility for studying the SEDs of
faint hosts.  From such SEDs the UV continuum flux can be determined and the
galaxy type can be inferred.  Previous investigations have shown that GRB
hosts have SEDs similar to starburst galaxies
\citep{sok01,goro02,goro03,lise03}, and their SFRs inferred from optical
methods are moderate $<10$~M\subsun \peryr\ 
\citep{fruchter99,bloom98b,djor98,djor01b}. Larger SFRs have been reported
based on spectroscopic measurements. The GRB 000418 host has an un-obscured
SFR of 55~M\subsun\,\peryr\, derived from the [\ion{O}{II}] line flux
\citep{bloom03}. All the optical methods for determining the SFRs are
affected by dust extinction in the hosts. Therefore, the optically inferred
SFRs represent lower limits to the true SFRs.  Radio and sub-mm data are much
less affected by dust extinction, and observations of GRB hosts indicate that
the unextincted SFRs can be as much as two orders of magnitude larger than
those derived from optical estimators \citep{berger01,berger02,tanvir04}.
However, not all hosts have very large SFRs; some have SFRs
$<200$~M\subsun\peryr\ suggested by radio observations \citep{vrees01b}.

The SFRs of individual GRB hosts published in the literature have been argued
to be comparable to those of other high redshift galaxies selected by optical
methods.  \cite{djor01c} found that the [\ion{O}{II}] equivalent widths of GRB
hosts are somewhat larger than that of field galaxies at similar redshifts as
GRBs. Likewise, \citet{fruc99b} found that a sample of three GRB hosts has
bluer colours on average than field galaxies in the Hubble Deep Field.

In this paper we present a statistical analysis of the properties of GRB hosts
as compared with other high redshift galaxies. We present the 10 GRB hosts
selected for the analysis in Sect.~\ref{host}, the derived photometric
redshifts in Sect.~\ref{zphot}, and the SED investigations in Sect.~\ref{sed}.
We estimate the SFRs of the 10 GRB hosts by computing the rest frame UV flux
in Sect.~\ref{sfr}. Since the absolute luminosity of the hosts vary by a large
factor we also analyse the specific SFRs normalised by the host luminosities.
Comparisons with properties of field galaxies selected from the Hubble Deep
Field are presented in Sect.~\ref{comp}. Our results are discussed and
summarised in Sect.~\ref{conc}.

We assume a flat cosmological model with $\Omega_m~=~0.3,
\Omega_{\Lambda}~=~0.7$ and $H_0~=~65$~km~s$^{-1}$~Mpc$^{-1}$. The
choice of parameters affects the luminosity distance of the galaxies,
and thereby the derived SFRs.


\section{GRB host galaxy sample}
\label{host}
Our own multi-colour imaging studies of GRB host galaxies have been presented
elsewhere \citep{goro02,goro03,lise03}. The present work is based on a
compilation of photometry already available in the literature (including our
own observational work). 

We imposed a magnitude limit to make sure that the hosts entering the sample
were bright enough to have fairly accurate multi-colour photometry in at least
5 optical and near-IR bands. This implies that the sample is limited by the
available multi-colour photometry from the literature. However, with a maximum
magnitude of $R=25.3$ the sample is magnitude-limited.  These criteria also
implied that no host of a GRB occurring after 2002 is included due to poor
multi-colour sampling.  We also required that a redshift for the host galaxy
or the afterglow be known.  Finally, we excluded a few GRB hosts which had
such complex morphologies that the resulting SEDs might be dominated by
different sub-components at different wavelengths (such as the
\object{GRB~980613}, \object{GRB~011121}, and \object{GRB~011211} hosts).
Having a complex morphology, the GRB~980613 host shows colour variations in
HST/STIS images of $m_{CL}-m_{LP}> 0.7$ between individual components
\citep{hjorth02}.  Similar colour variations was found in ground based
observations by \citet{djorgov03}.  Such variations in colours make any
detailed analysis of the overall SED subject to great uncertainty in terms of
the derived extinction and age.  Also the GRB~011121 and GRB~011211 hosts have
complex surroundings, but not much is presently known about the colours of the
hosts themselves \citep{garnavich03,greiner03,jakobsson03}.  We therefore
chose to exclude these hosts, but note that these systems represent likely
mergers which may show significant star formation. Images of the various hosts
are presented elsewhere \citep{bloom00,djorgov03,castro03}.

To summarize, our selection criteria are:
\begin{itemize}
\item  Detection in 5 or more filters
\item $R<25.3$
\item Known redshift
\item Not very complex morphology
\end{itemize}

These criteria limited our investigations to 10 GRB hosts in the redshift
range $0.433<z<2.037$ with a mean and median redshifts of $z=0.97$ and
$z=0.85$, somewhat smaller than those derived from the 35 GRB redshifts
measured to date ($z=1.43$ and $z=1.10$) \footnote{GRB redshifts can be found
  at Jochen Greiner's web-pages: \tt
  http://www.mpe.mpg.de/$\sim$jcg/grbgen.html}.  The sample is presented in
Table~\ref{tab:allmag}.

Some of the host magnitudes were obtained a few weeks after the burst when the
optical afterglow could still contaminate the observed flux. In these cases
the host magnitudes are derived from fits to the light curves of the
afterglows.  Since the light curve of the afterglow can be described by a
power law, the total flux is given by
\(f(t)=f_0~\times~t^{-\alpha}+f_{\mathrm{host}}\), where the first term
characterizes the fading afterglow. If the light curve is well sampled, the
flux of the host $f_{\mathrm{host}}$ can be estimated.  For example, the GRB
980703 host magnitudes were derived this way in \citet{vrees99}. Data obtained
more than one year after this particular burst gave magnitudes which are
consistent with those reported in Table~\ref{tab:allmag} \citep{holland01}.
As another example, the expected $B$ band magnitude of the GRB~010921
afterglow would be 3 magnitudes fainter than the host magnitude reported in
\citet{price02} at the time of the observations 21 days after the burst.

Ideally, the magnitudes of a host should be derived using one consistent
photometric technique for all filters. For example, in the case of aperture
photometry the magnitudes should be derived using the same aperture. We can
not be sure that this is the case for the magnitudes given in
Table~\ref{tab:allmag}. In the cases where the hosts are more extended than
point sources the effect should be negligible as long as the authors have
applied a large enough aperture for deriving the host magnitudes.

Furthermore, one should note that we have restricted ourselves to analysing
only the bright end of the luminosity function since it is easier to perform
multiband observations of the brightest hosts.  Only little is known about the
nature of the fainter host galaxies \citep{berger02b,jaunsen03,hjorth03b}.
 
\begin{table*}
\begin{center}
\begin{tabular}{llll}
\hline \hline
\noalign{\smallskip}
Host  & Filter (Mag) & Telescope & Reference\\
\noalign{\smallskip}
\hline
\noalign{\smallskip}
\object{GRB 970228} & $B>26.08$         &VLT       &\cite{sok01}\\
          & $V=25.77\pm0.2$   &HST/STIS  & \cite{gal00b}\\
          & $R_c=25.22\pm0.2$ &HST/STIS  &\\
          & $I=24.4\pm0.2$    &  HST/WFPC2  &\cite{fruchter99}\\
          & $H=23.2\pm$0.3 &HST/NICMOS2\\
          & $K=22.6\pm$0.2    & NIRC/Keck I  &\cite{chary01}\\
\noalign{\smallskip}
\object{GRB 970508} & $B=25.89\pm$0.19    &  BTA     &\cite{sok01}\\
 &$V=25.34\pm$0.22 &BTA\\
 &$R_c=25.06\pm0.17$&BTA\\
 &$I_c=24.11\pm0.25$&BTA\\
&$K=22.7\pm0.2$&   Keck I/NIRC&  \cite{chary01}\\
\noalign{\smallskip}
\object{GRB 980703} &  $B=23.40\pm0.12$ &BTA&\cite{sok01}\\
 & $V=23.04\pm$0.08 $^{\dagger}$ &   &\cite{vrees99}\\
&  $R=22.58\pm$0.06\\
&  $I=21.95\pm$0.25\\
& $J=20.87\pm$0.11\\
& $H=20.27\pm$0.19\\
& $K=19.62\pm$0.12\\
\noalign{\smallskip}
\object{GRB 990123} & $U=23.6\pm$0.15 &   & \cite{castro99}\\
& $B=24.23\pm$0.1 $^{\dagger}$ & \\
& $V=24.20\pm$0.15\\
& $R=23.77\pm$0.1\\
& $I=23.65\pm$0.15\\
& $K=21.7\pm$0.3  & Keck I/NIRC& \cite{chary01}\\
\noalign{\smallskip}
\object{GRB 990712} &  $U$ =  23.12$\pm$0.05 &ESO-3.6m& \cite{lise03}\\
          & $B$ =  23.36$\pm$0.09& DK-1.5m/DFOSC&\\ 
          & $V$=   22.39$\pm$0.03 & DK-1.5m/DFOSC&\\ 
          & $R$=   21.84$\pm$0.02 & DK-1.5m/DFOSC&\\ 
          & $I$=   21.41$\pm$0.03& DK-1.5m/DFOSC&\\ 
          & $J$=   20.81$\pm$0.17& NTT/SOFI\\
          & $H$=   20.25$\pm$0.19& NTT/SOFI\\
          & $K\!s$=20.05$\pm$0.1&  & \cite{lefloch03}\\
\noalign{\smallskip}
\object{GRB 991208} & $B=25.19\pm$0.17 $^{\dagger}$ & & \cite{cas01}\\
            &   $V=24.55\pm$0.16\\
            &   $R=24.26\pm$0.15\\
            &   $I=23.3\pm$0.2\\
            &   $K=21.7 \pm$0.2 & NIRC/Keck I &\cite{chary01}\\
\noalign{\smallskip}
\hline
\end{tabular}
\caption[]{Magnitudes in the Vega system in various filters for the selected sample of 10
  GRB hosts taken from the literature. The magnitudes are not corrected for
  Galactic extinction. $^{\dagger}$ indicates that the host magnitudes were
  derived from power law fits to the afterglow light curves.}
\label{tab:allmag}
\end{center}
\end{table*}
\addtocounter{table}{-1}
\begin{table*}
\begin{center}
\begin{tabular}{llll}
\hline \hline
\noalign{\smallskip}
Host  & Filter (Mag) & Telescope & Reference\\
\noalign{\smallskip}
\hline
\noalign{\smallskip}
\object{GRB 000210} & $U=23.54\pm0.13$ & ESO-3.6m/EFOSC2 & \cite{goro02}\\
        &   $B=24.40\pm0.13$ & ESO-3.6m/EFOSC2&\\
        &   $V=24.22\pm$ 0.08 & ESO-3.6m/EFOSC2 \\
        &   $R=23.46\pm$ 0.10 &VLT/FORS1 &\cite{piro02}\\
        &   $I=22.49\pm$ 0.12 & ESO-3.6m/EFOSC2 &\cite{goro02}\\
        &   $Z=22.83\pm0.28$& DK-1.5m/DFOSC\\
        &   $J\!s=21.98\pm$ 0.10 &VLT/ISAAC\\
        &   $H=21.51\pm$ 0.23 &NTT/SOFI&\\
        &   $K\!s=20.94\pm$ 0.14 &VLT/ISAAC\\
\noalign{\smallskip}
\object{GRB 000418}  & $U=23.54\pm0.3$ & ESO-3.6m/EFOSC2 &\cite{goro03}\\ 
            &   $B=24.07\pm$0.05 & NOT/ALFOSC \\
            &   $V=23.80\pm$0.06 &NOT/ALFOSC \\
            &   $R=23.36\pm$0.05&NOT/ALFOSC\\
            &   $I=22.79\pm$0.05 &NOT/ALFOSC\\
            &   $Z=22.46\pm0.1$ &  NOT/ALFOSC\\
            &   $J\!s=22.27\pm0.1$ & VLT/ISAAC\\
           &   $K\!s=21.19\pm 0.3$ & VLT/ISAAC \\
\noalign{\smallskip}
\object{GRB 000926}  & $B  = 25.49\pm$0.33 $^{\dagger}$ & & \cite{castro03}\\
            & $V  = 25.08\pm$0.06  \\
            & $Rc = 24.83\pm$0.07  \\
            & $Ic = 24.59\pm$0.01   \\
            & $J_{AB}  = 24.1^{+0.7}_{-0.4}$ & & J. Fynbo (priv. comm.) \\
\noalign{\smallskip}
\object{GRB 010921} & $B=23.42\pm0.08$ &  & \cite{price02}\\
            & $V  = 22.32\pm$0.06  \\
            & $Rc = 21.93\pm$0.09  & & \cite{park02} \\
            & $Ic = 21.05\pm$0.08   \\
            &   $J=20.34\pm0.02$ & Keck I/NIRC & \cite{price02}\\
            &   $H=19.75\pm0.04$ & Keck I/NIRC\\
            &   $K\!s=19.07\pm 0.04$ & Keck I/NIRC \\
\noalign{\smallskip}
\hline
\end{tabular}
\caption[]{Continued. }
\end{center}
\end{table*}

\section{Photometric redshifts}
\label{zphot}
The magnitudes in Table \ref{tab:allmag} were used for comparison with
theoretical galaxy template spectra from \cite{bru93}. This was done using the
program HyperZ\footnote{\tt http://webast.ast.obs-mip.fr/hyperz/} developed by
\citet{bolzo00}.
 
The templates which are used to fit the GRB host magnitudes consist of
elliptical, different types of spiral galaxies, irregular, and starburst
spectra at various ages having different star-formation histories. The SFRs
decrease with time, such that SFR~$\propto \exp({-t/\tau})$, where the
characteristic timescale, $\tau$ increases along the Hubble sequence.  The
starburst template is created as an instantaneous burst of star formation
($\tau \rightarrow 0$), which instantly uses up all available gas, while an
irregular Im template has a constant SFR ($\tau \rightarrow \infty$). We used
the \citet{miller79} initial mass function (IMF) for calculating the
templates, as well as a \citet{salpeter55} IMF with stellar masses between 0.1
and 125~M\subsun. The Miller \& Scalo IMF produces fewer massive stars
compared to a Salpeter IMF and is flatter below 1~M\subsun. The metallicities
of the templates were solar, $Z=0.02$.

Applying an additional extinction term to the templates, the fitting allows an
estimate of the type of galaxy, age, and the intrinsic extinction ($A_V$) for
the galaxies.  In the fits we used the extinction curve estimated for
starburst galaxies \citep{calz00}. We also analysed the SEDs using other
extinction curves, e.g. from the Milky Way \citep{seaton79}, the Large
Magellanic Cloud \citep{fitz86}, and the Small Magellanic Cloud
\citep{prevot84}. We found that the derived ages for the dominant population
of stars and extinctions did not depend on the chosen extinction curve. The
reason is that actual values of the extinctions are quite small ($A_V<1$),
thus the various extinction curves produce minor differences, as quantified in
Sect.~\ref{sect:age_ext}.

A direct application of HyperZ results in the photometric redshifts listed in
Table~\ref{tab:hypsed}.  In most cases these are consistent with the
spectroscopic ones. The mean value and standard deviation of
($z_{\mathrm{phot}}-z_{\mathrm{spec}}$) is --0.04 and 0.21, respectively. Only
for the GRB 990123 host the redshift estimate is inaccurate ($\Delta z>0.3$),
but taking into account the uncertainty of the estimate the difference is only
1.6$\sigma$. A reason for the relatively large discrepancy could be that the
Balmer jump is not well sampled. Indeed, very accurate photometric redshifts
can be determined if the photometric uncertainties are small and the Balmer
jump is well sampled as is the case for the GRB~000418, GRB~000210, and
GRB~990712 hosts.
  
At higher redshifts the broad band filters cover a narrower wavelength range
of the rest frame spectrum due to the factor $(1+z)$ accounting for the
cosmological expansion.  Therefore, the accuracy of photometric redshift
estimates is expected to decrease with increasing redshift. A different
measure of the accuracy can be obtained by calculating the expression \(\Delta
z~=~(z_{\mathrm{phot}}-z_{\mathrm{spec}})/(1+z_{\mathrm{spec}})\) for the
sample of GRB hosts. We find a mean value of 0.015 and a standard deviation of
0.16 using this expression, showing that it is possible to determine
photometric redshifts accurately for GRB hosts.  For comparison,
\citet{fernandez01} derived more accurate photometric redshifts for galaxies
observed through 7 bands in the Hubble Deep Field, having a standard deviation
of 0.065, which is likely due to smaller photometric uncertainties for their
galaxies. While the average number of bands of observations of the GRB host
sample is 6.6, 3 of the hosts have observations in 5 filters only.
Additionally, the standard broad band filters used for ground based photometry
are not optimally tuned to find photometric redshifts. We find no outliers for
the estimation of photometric redshifts, indicating that this technique is
robust for estimating redshifts of GRB hosts.

\section{Spectral energy distributions}
\label{sed}

While HyperZ was written for obtaining photometric redshifts of galaxies in
large surveys, it also serves the purpose of finding the best matching
theoretical galaxy template for a given set of broad-band observations.  In
the remainder of this work we shall fix the redshifts of the GRB hosts to the
values given by the spectroscopic measurements. This was done in order to
optimise the estimates of other output parameters, as explained below.

\begin{figure*}
\centering
\vspace*{-2cm}
\resizebox{\hsize}{25cm}{\includegraphics[bb= 110 100 526 900, clip]{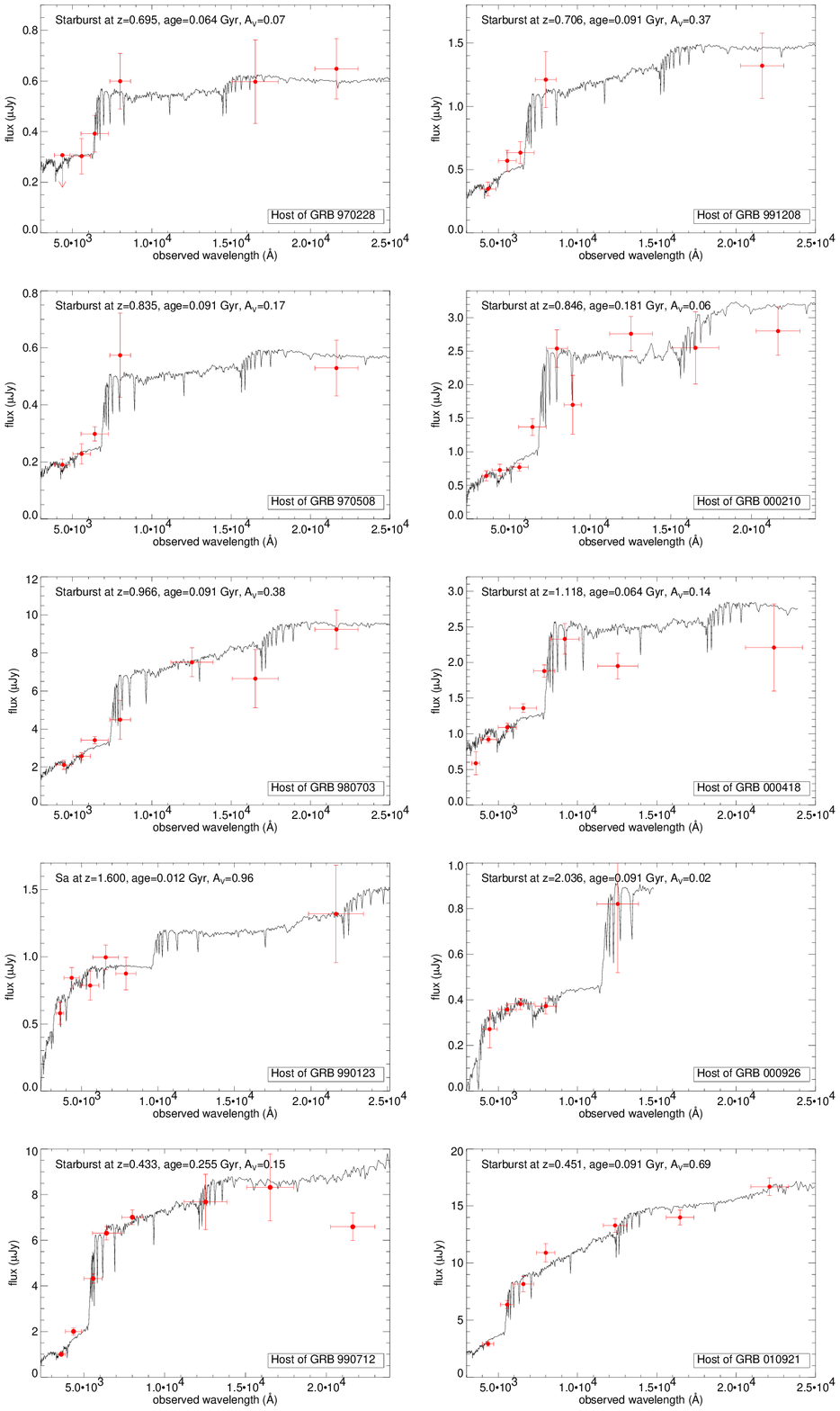}}
\caption{Best fits of the 10 GRB host SEDs fitted to synthetic spectra based
  on a Salpeter IMF as calculated using HyperZ. Redshifts, type of the
  template, and the extinction are given as inserts in the plots. The solid
  circles show the available photometry for each host (see
  Table~\ref{tab:allmag}) after correcting for Galactic extinction.  The
  associated horizontal error bars denote the FWHM of the filters. }
\label{fig:hypplots}
\end{figure*}

The best fit is obtained by minimizing the expression
\begin{equation}
\chi^2=\sum_i \Big( \frac{F_{\mathrm{host},i} - k \times F_{\mathrm{temp},i}}{\sigma(F_{\mathrm{host},i})} \Big)^2
\label{eq:chi}
\end{equation}
where the sum is to be taken over all filters, $i$. $F_{\mathrm{host}}$ is the
flux density of the host in the filter $i$, $\sigma(F_{\mathrm{host},i})$ is
the associated error, and $k$ is a normalization constant.
$F_{\mathrm{temp},i}$ is the flux of the template in the filter $i$, which is
calculated using the throughput for the given filter and instrument. HyperZ
provides a data set of throughputs for many instruments. For the instruments
where we had no knowledge of the throughput (as for example for the Russian
6m-BTA telescope), we simply used the throughput of the given filter used for
the observation, i.e. without convolving with the quantum efficiency of the
CCD. We tested to see if it had any significance using e.g. the VLT throughput
curve for the given filter and found the effect to be negligible. Since the
photometric errors are large in these cases, the results for the best fit
template did not change.  In all cases, only templates from one single burst
of star formation were used. This is a simplification since more than one
population of stars may be present in the hosts \citep{sok01,lise03}.
Hereafter the derived stellar properties will refer exclusively to the
dominant population.

Detailed analyses of the SEDs of the host galaxies of GRB~000210, GRB~000418,
and GRB~990712 are presented by \citet{goro02}, \citet{goro03}, and
\citet{lise03}, respectively. Similar thorough individual SED analyses are
beyond the scope of this paper.

Results from the SED fittings are given in Table~\ref{tab:hypsed}. Column 2
gives the measured spectroscopic redshift, which was held fixed while running
HyperZ. For comparison the unconstrained photometric redshifts are listed in
column 3. In column 4 the best fit templates are given, in column 5 the age of
the template, column 6 gives the extinction, and column 7 lists the reduced
$\chi^2$ per degree of freedom (d.o.f.) for the best fits.  All these values
are derived using templates constructed using a Miller \& Scalo IMF.  Column 8
and 9 list the extinction and $\chi^2$/d.o.f. for fits using a Salpeter IMF.
The age and galaxy type do not change choosing a different IMF, and all the
observed host SEDs are well fit by starburst templates or young star-forming
galaxy types.

Generally, the ages of the dominant population of stars are smaller than
0.2~Gyr for all GRB hosts, and the extinctions found for the systems are
relatively small: $0<A_V<2$.  Plots of the best fit templates for each host
are shown in Fig.~\ref{fig:hypplots}.

\subsection{Ages and extinctions}
\label{sect:age_ext}
To estimate errors on ages and extinctions we analysed the values of $\chi^2$
for SED fits of various templates while varying the extinction.  We derived
errors of the extinction and age on the basis on which fits gave an increase
in the $\chi^2$ by $\Delta(\chi^2)=1$. The host of GRB 000418 and GRB 980703
are treated separately below.

Using different extinction laws resulted in differences in $A_V$ smaller than
0.1, and furthermore, the estimated ages remained constant for any applied
extinction law.

\begin{table*}
\begin{tabular}{ccccccc| cc}
\\
\hline \hline
\noalign{\smallskip}
&&&Miller \& Scalo  IMF & &&& Salpeter IMF\\
\hline
\noalign{\smallskip}
Host &  $z_{\mathrm{spec}}$ & $z_{\mathrm{phot}}$ & spectrum & age (Gyr) & $A_V$ & $\chi^2$/d.o.f. & $A_V$ & $\chi^2$/d.o.f.\\
\noalign{\smallskip}
\hline
\noalign{\smallskip}
\object{GRB 970228} &0.695 &$0.78^{+0.24}_{-0.12}$  & burst & 0.064$\pm$0.012
& 0.12$^{+0.18}_{-0.12}$ & 0.228 & 0.07$^{+0.16}_{-0.07}$ & 0.239\\
\object{GRB 970508} &0.835 &$0.87^{+0.07}_{-0.12}$  & burst & 0.091$\pm$0.090 & 0.17$^{+0.11}_{-0.17}$ & 0.146 & 0.17$^{+0.11}_{-0.17}$ & 0.082\\
\object{GRB 980703} &0.966 &$0.82^{+0.16}_{-0.18}$  & burst & 0.006$\pm$0.122 & 1.60$^{+0.05}_{-1.27}$ & 0.683 & 0.38$^{+0.25}_{-0.24}$ & 0.831\\
\object{GRB 990123} &1.600 &$2.18^{+0.31}_{-0.40}$  & Sa    & 0.012$\pm$0.006 & 0.90$^{+0.17}_{-0.20}$ & 0.823 & 0.96$^{+0.05}_{-0.23}$ & 0.773\\
\object{GRB 990712} &0.433 &$0.42^{+0.003}_{-0.01}$ & burst & 0.255 &
0.15$^{+0.04}_{-0.05}$ & 2.815 & 0.15$^{+0.06}_{-0.05}$ & 0.959\\
\object{GRB 991208} &0.706 &$0.78^{+0.02}_{-0.08}$  & burst & 0.091$\pm$0.090 & 0.35$^{+0.18}_{-0.35}$ & 0.621 & 0.37$^{+0.16}_{-0.30}$ & 0.613\\
\object{GRB 000210} &0.846 &$0.84^{+0.05}_{-0.04}$  & burst & 0.181 & 0.02$^{+0.06}_{-0.02}$ & 1.225 & 0.06$^{+0.08}_{-0.06}$ & 1.105\\
\object{GRB 000418} &1.118 &$1.00^{+0.02}_{-0.003}$ & burst & 0.064$\pm$0.027 & 0.12$^{+1.50}_{-0.12}$ & 2.448 & 0.14$^{+1.40}_{-0.14}$ & 2.266\\
\object{GRB 000926} &2.037 &$2.09^{+0.54}_{-0.59}$  & burst & 0.091$\pm$0.060 & 0.00$^{+0.44}$ & 0.020 & 0.02$^{+0.43}_{-0.02}$ & 0.025\\
\object{GRB 010921} &0.451 &$0.37^{+0.02}_{-0.02}$  & burst & 0.091$\pm$0.090 & 0.68$^{+0.05}_{-0.02}$ & 2.361 & 0.69$^{+0.05}_{-0.27}$ & 2.353\\
\noalign{\smallskip}
\hline
\end{tabular}
\caption{Results from the best fits from HyperZ.  Column 2 gives the
spectroscopic redshift of the hosts, which was held fixed while
running HyperZ. Column 3 lists the unconstrained photometric
redshifts  and the associated 68\% confidence levels.
 The best fit template is given in column 4, and the
corresponding ages and internal extinctions are given in column 5 and
6, respectively, using a Miller \& Scalo IMF. Column 8 and 9 list the
results of the extinction and $\chi^2$/d.o.f. for a Salpeter IMF.}
\label{tab:hypsed}
\end{table*}

Since GRBs are thought to be produced by the most massive stars, ages much
larger than 10 Myr of the burst population are inconsistent with this
hypothesis. However, we can not exclude the possibility that two separate
bursts of star formation would be able to reproduce the SEDs of the hosts. The
burst template model uses up all gas in the first burst, while a more reliable
model would likely have $\tau>0$. We have fit all host magnitudes to Im
models, which generally produced acceptable fits ($\chi^2$/d.o.f. $<2$).
However, this was not the case for the GRB~990712, GRB~000210, and GRB~000418
hosts, which gave $\chi^2/$d.o.f.$~=~14$, 24, and 6, respectively. These hosts
are the only ones which have extensive multiband photometric measurements, and
their SEDs are well constrained. This implies that a continuous SFR is not in
agreement with the observed SED. The large errors for the photometry of the
other hosts make their SEDs less well constrained.

A special comment is needed for the host of GRB 000418. In \citet{goro03}, the
final reported results for the SED fits are consistent with those reported
here in Table~\ref{tab:hypsed}. However, the actual best fit having the
minimum $\chi^2$/d.o.f. for a Salpeter IMF and a Calzetti extinction law, is
obtained with a 0.004 Gyr old starburst template with an extinction of
$A_V=1.38$ (see Table~3 in \cite{goro03}). We will return to the implication
of this uncertainty in Sect.~\ref{sfr_comp}.

As seen in Table~\ref{tab:hypsed} the estimated extinction for the GRB 980703
host changes by a large factor for the two applied IMFs. However, the lower
limit on the extinction derived for the Miller \& Scalo IMF fit is consistent
with that derived for the Salpeter IMF fit.

\subsection{Ages and metallicities}
  
There is a well known degeneracy between the age of a stellar population and
the metallicity for a given SED. In order to quantify how much this degeneracy
affects our results we fitted the SEDs to starburst templates, created from a
Salpeter IMF, with metallicities of 1, 0.4, and 0.2 times solar, respectively,
using the GALAXEV library of evolutionary stellar population synthesis models
\citep{bruzual03}.  The ages of the templates were between 5 Myr and 200 Myr.
While keeping the extinction values fixed to those obtained in
Table~\ref{tab:hypsed} the best fit ages for the different metallicities are
listed in Table~\ref{tab:age_test}. Compared to the ages found by HyperZ in
Table~\ref{tab:hypsed}, there are only small differences which are likely due
to the differences for the input templates. Moreover, there is a general
agreement between the best fit ages for the various metallicities. Because of
this result we estimate that for the GRB hosts the age-metallicity degeneracy
produces small systematic errors.
  
Alternatively, as shown from the analysis of the GRB~000418 host in
\citet{goro03}, the metallicity is not strongly constrained in the case where
the SED is analysed through broad band magnitudes.
  
For the median redshift $z\approx1$ of the GRB hosts our SED analysis covers
the rest-frame far-UV to rest-frame $J$ bands.  Using SED analyses of star
clusters in the nearby \object{NGC~3310} starburst galaxy \citet{degrijs03}
find that the ages are well reproduced when UV-near-IR pass bands are
included, which supports our finding of a consistent best fit age.


\section{Star-formation rates}
\label{sfr}
Since mostly young stars contribute to the UV flux in a galaxy there is a
relation between the rest frame UV continuum flux of a galaxy and the
unobscured SFR.  One can estimate the SFR from the rest frame flux at
2800~{\AA} using the relation in \citet{kennicutt98}
\begin{equation}
\textrm{SFR}\,(M_{\odot} \peryr) = 1.4 \times 10^{-28}  L_{\nu,\mathrm{UV}} \quad [\textrm{erg}~\textrm{s}^{-1}~\textrm{Hz}^{-1}].
\label{eq:sfruv}
\end{equation}
This method is appropriate for obtaining the SFR as long as there is a
continuous formation of stars. It is a good estimator for ages larger than
10$^8$ years, but shows limitations for ages smaller than 10$^7$ years. Having
found that the ages of some of the GRB hosts could be smaller than 10$^{7}$
years, the relation (\ref{eq:sfruv}) is not always applicable. For younger
populations the constant linking the SFR with the luminosity is significantly
lower, yielding a smaller SFR for a given flux. On the other hand, dust is
expected to be present in star-forming regions in which case the observed flux
must be corrected for dust extinction. These two effects have an opposite
impact on the derived SFRs.

The method applied to calculate the SFRs from the observed broad band
magnitudes was as follows. First the magnitudes were corrected for Galactic
extinction using the dust maps of \citet{schlegel98}. Magnitude offsets
calculated by HyperZ from standard filters to the AB system were added.  The
AB system is defined as \(m_{\mathrm{AB}}=-2.5 \log f_{\nu}-48.6\) where
$f_{\nu}$ is the flux density measured in erg~cm$^{-2}$~s$^{-1}$~Hz$^{-1}$.
Broad band magnitudes were converted to flux units (in $\mu$Jy) using the
expression \(f_{\nu}= 10^{-0.4(m_{\mathrm{AB}}-23.9)}\).  The fluxes at the
observed wavelengths 2800$(1+z)$~{\AA} were estimated assuming power law
spectra, \(f_{\nu}=f_0 (\nu/\nu_0)^{\beta}\) for the hosts, between the two
filters bracketing the observed wavelength at 2800$(1+z)$~{\AA}.

The calculated SFRs for the 10 GRB hosts are listed in
Table~\ref{tab:sfrground}. Knowing the extinction of the hosts from the SED
analyses, one can correct the SFRs for the effects of extinction. Using the
extinction curve of \cite{calz00} we derive the unextincted SFRs given in
column 6.

The errors of the SFRs in Table~\ref{tab:sfrground} are due to the photometric
errors used for interpolation which translates into errors of the 2800~{\AA}
fluxes. The intrinsic scatter in the calibration converting UV flux into SFR
is of the order of 30\% \citep{kennicutt98}. This uncertainty is not included
in the quoted SFRs.

\begin{table}
\begin{tabular}{lllll}
\hline \hline
\noalign{\smallskip}
Host & $A_V$ (fixed) &   Z=0.02  & Z=0.008 & Z=0.004 \\
\noalign{\smallskip}
\hline
\noalign{\smallskip}
GRB 970228  & 0.07 & 100 & 50  & 50 \\
GRB 970508  & 0.17 & 100 & 100 & 100\\
GRB 980703  & 0.38 & 100 & 100 & 100\\
GRB 990123  & 0.96 & 5   & 5   & 5 \\ 
GRB 990712  & 0.15 & 200 & 200 & 200\\
GRB 991208  & 0.37 & 100 & 100 & 100\\
GRB 000210  & 0.06 & 200 & 200 & 200\\
GRB 000418  & 0.14 & 70  & 100 & 100\\
GRB 000926  & 0.02 & 100 & 100 & 100\\
GRB 010921  & 0.69 & 100 & 100 & 100 \\
\noalign{\smallskip}
\hline 
\end{tabular}
\caption{Ages in Myr derived for fits to various metallicities. In addition to
  the redshifts, also extinction values are held fixed.}
\label{tab:age_test}
\end{table}

For comparison, SFRs derived from spectroscopic measurements of the
[\ion{O}{II}] line flux are listed in column 7 in Table~\ref{tab:sfrground}.
Apart from the GRB~000418 and GRB~991208 hosts the agreement between the
unextincted, UV based and [\ion{O}{II}] based SFRs is rather good which
suggests that extinction does not play a major role.  Values of the SFR
derived either from sub-mm, or radio observations are also listed in
Table~\ref{tab:sfrground}.  These measurements generally show larger values,
which may indicate very obscured components with no (or faint) optical
emission within the galaxies.

\begin{table*}
\begin{tabular}{ccccccll}
\\
\hline \hline
\noalign{\smallskip}
 Host  & $z$ & $d_L$ & UV flux  & SFR  &
 unextincted SFR   & [\ion{O}{II}] based SFR  (refs.) & radio/sub-mm SFR\\
     & & (cm) & ($\mu$Jy) & (M$_{\odot}$ yr$^{-1}$)& (M$_{\odot}$ yr$^{-1}$) &
     (M$_{\odot}$ yr$^{-1}$) & (M$_{\odot}$ yr$^{-1}$)\\
\noalign{\smallskip}
\hline 
\noalign{\smallskip}
\object{GRB 970228} & 0.695  &1.40$\times 10^{28}$   & 0.34$\pm$0.16  & 
 0.70$\pm$0.32      & 0.78$\pm$0.36  &  0.76 (1) & $<$335 (8) \\
\object{GRB 970508} & 0.835  &1.76$\times 10^{28}$   & 0.28$\pm$0.15  & 
 0.83$\pm$0.45      & 1.10$\pm$0.60  &  1.4 (2) & $<$380 (8)\\
\object{GRB 980703} & 0.966  &2.10$\times 10^{28}$   & 3.20$\pm$0.08  & 
 12.7$\pm$0.32      & 23.8$\pm$0.60  & 20 (3) & 180$\pm$25 (8)\\
\object{GRB 990123} & 1.600   &3.93$\times 10^{28}$   & 0.55$\pm$0.16 & 
 5.71$\pm$1.69      & 28.0$\pm$8.29  & \\
\object{GRB 990712} & 0.433  &7.95$\times 10^{27}$   & 1.65$\pm$0.04  & 
 1.28$\pm$0.03      & 1.64$\pm$0.04  & 2.12$\pm$0.60 (4) & $<$100 (9)\\
\object{GRB 991208} & 0.706 &1.43$\times 10^{28}$   & 0.40$\pm$0.13  & 
 0.83$\pm$0.28      & 1.54$\pm$0.52 & 4.8$\pm$0.2 (5) & 70$\pm$30 (8) \\
\object{GRB 000210} & 0.846   &1.77$\times 10^{28}$   & 0.79$\pm$0.07  & 
 2.37$\pm$0.20      & 2.62$\pm$0.28 & 3 (6) & 90$\pm$45 (8)\\
\object{GRB 000418} & 1.118  &2.52$\times 10^{28}$   & 1.33$\pm$0.04  & 
 7.02$\pm$0.22      & 8.85$\pm$0.22 & 55 (7) & 330$\pm$75 (8)\\
\object{GRB 000926} & 2.037 &5.28$\times 10^{28}$   & 0.50$\pm$0.26  & 
 8.09$\pm$4.16      & 8.37$\pm$4.30 &   & 820$\pm$340 (8)\\
\object{GRB 010921} & 0.451 & 8.32$\times 10^{27}$ & 2.15$\pm$0.08  &  
 1.81$\pm$0.07      & 5.67$\pm$0.22 & \\
\noalign{\smallskip}
\hline
\end{tabular}
\caption{SFRs calculated from ground based GRB host
  observations. Column 2 lists the spectroscopic redshifts and column
  3 the corresponding luminosity distances. Column 4 and 5 list the
  inferred UV flux and SFRs respectively, and column 6 the SFRs
  corrected for the internal extinction from Table~\ref{tab:hypsed}.   The intrinsic scatter of 30\% for the UV to SFR calibration  has not been
  included in the reported errors. The last two columns list the SFRs derived 
  from spectroscopic measurements of the [\ion{O}{II}] lines and sub-mm/radio 
  observations taken from following references: (1) \citet{bloom01}, (2)
  \citet{bloom98b}, (3) \citet{djor98}, (4) \citet{hjorth00,hjorth00b}, (5)
  \citet{cas01}, (6) \citet{piro02}, (7) \citet{bloom03}, (8)
  \citet{berger02}, (9) \citet{vrees01b}. Note that these SFRs are sometimes
  derived using a different cosmology than adopted in this paper which will
  cause a small change in the derived SFR.}

\label{tab:sfrground}
\end{table*}

\begin{figure}
\centering
\resizebox{\hsize}{!}{\includegraphics[bb=35 0 585 405,clip]{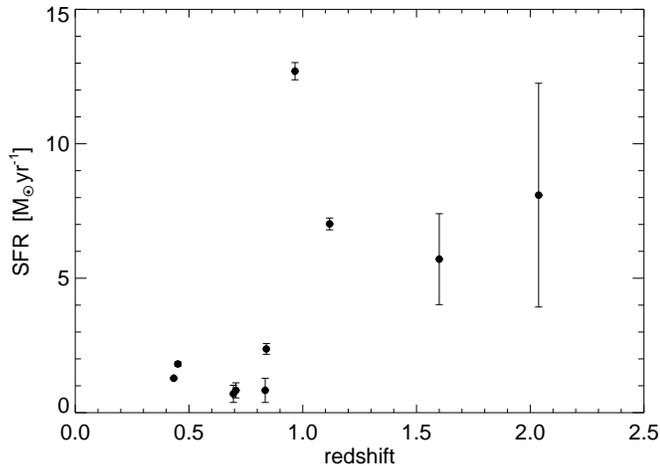}}
\caption{Star-formation rates of the 10 GRB hosts as a function of
  redshifts. The SFRs have not been corrected for the effect of host galaxy
  extinction.  A trend of larger SFRs for the hosts at larger redshifts is
  very likely a selection effect.}
\label{fig:sfrgrb}
\end{figure}

One immediately sees from Table~\ref{tab:sfrground} that the SFRs of the hosts
are moderate, in line with the conclusions of
\citet{fruchter99,djorgov03,djor01b,cas01}.  In Fig.~\ref{fig:sfrgrb} the
calculated SFRs are plotted as a function of redshift. Since the faint,
high-redshift hosts do not have multiband observations, and therefore are not
studied here, the trend for larger SFRs at high redshifts may be caused by the
selection of intrinsically bright hosts.

For the GRB 000926 host at $z=2.037$ and the GRB~990123 host at $z=1.600$ we
use the observed $B$ and $U$ band magnitudes, respectively, as a rough measure
of their continuum flux at 1500~{\AA} in the rest frame to get an independent
estimate of the SFR (uncorrected for extinction).  Using the relation in
\citet{madau98}
\begin{equation}
\textrm{SFR~(M}\subsun \peryr)=1.3\times10^{-28} L_{\nu} \quad \textrm{[erg~s$^{-1}$~Hz$^{-1}$]}
\label{eq:sfr2}
\end{equation}
we find SFR = 4.3$\pm$2.0 and 7.4$\pm$3.0 M\subsun \peryr\ not corrected for
internal extinction for the two hosts, respectively. In principle, the SFR
derived from the continuum at smaller wavelengths could be used to constrain
the intrinsic extinction, but in practice this is difficult given the
photometric uncertainties. Indeed, one measurement appears to give a smaller
SFR based on the 1500~{\AA} calibration compared to the 2800~{\AA}
calibration, while the other is slightly larger.  However, in both cases the
SFRs are consistent within 1$\sigma$ errors.

\subsection{Specific SFRs}
Even though the SFRs appear small, they show variations of more than a factor
of 10 between the individual hosts. A more informative measure of the
star-formation activity in the galaxies may be the SFR per unit luminosity.
The rest-frame $B$ band luminosities are calculated from the best fit (rest
frame) spectra by convolving with the $B$ band filter transmission. This way,
all information from the broad band observations is used, and no K-correction
is involved. Using the cosmological model we can estimate the absolute $B$
band magnitude and the corresponding luminosity of the galaxy, $L$, given in
Table~\ref{tab:sfrnorm}.

It is evident that some of the hosts are less luminous than an $M^*$ galaxy,
i.e., the magnitude of a galaxy at the break in the Schechter luminosity
function \citep{schechter}. A similar conclusion based on colours of a sample
of GRB hosts was reached by \citet{lefloch03}.  We adopt $M^*=-21$ which is
typically inferred for field galaxies. At higher redshifts this magnitude is
reported to vary for blue galaxies \citep{lilly95}, but since all galaxies are
treated identically here the actual value of $M^*$ will just introduce a
systematic change of the specific SFRs.

The SFRs from Table \ref{tab:sfrground} were divided by the quantity $L/L^*$
in order to calculate specific SFRs presented in Table~\ref{tab:sfrnorm}.  The
specific SFRs (not corrected for extinction) vary by a factor of $\sim$2, as
shown in the upper panel of Fig.~\ref{fig:normsfrhdf}.  Choosing a different
magnitude for an $M^*$ galaxy does not change this result.  The distribution
of specific SFRs has a mean of 9.7 M\subsun \peryr\ ($L/L^*$)$^{-1}$ and a
standard deviation of 2.1. Since the SFR is a measure of the flux in the UV
rest frame, and the luminosity is a measure of the rest frame flux at
$\sim$4400~{\AA}, the specific SFR is simply characterizing the slope of the
spectrum for each host.

\begin{table}
\begin{tabular}{lllll}
\\
\hline \hline
\noalign{\smallskip}
 Host & $z$ & $M_B$ & $L/L^*$ & specific SFR\\
&&&& (M\subsun \peryr\, ($L/L^*$)$^{-1}$) \\
\noalign{\smallskip}
\hline 
\noalign{\smallskip}
\object{GRB 970228} & 0.695 & --18.08 & 0.07 & 10.3$\pm$4.71 \\
\object{GRB 970508} & 0.835 & --18.14 & 0.07 & 11.6$\pm$6.27 \\
\object{GRB 980703} & 0.966 & --21.39 & 1.43 & 8.87$\pm$0.22 \\
\object{GRB 990123} & 1.600 & --21.05 & 1.05 & 5.45$\pm$1.61 \\
\object{GRB 990712} & 0.433 & --18.56 & 0.11 & 12.1$\pm$0.28 \\
\object{GRB 991208} & 0.706 & --18.48 & 0.10 & 8.45$\pm$2.85 \\
\object{GRB 000210} & 0.846 & --19.36 & 0.22 & 10.7$\pm$0.91 \\
\object{GRB 000418} & 1.118 & --20.41 & 0.58 & 12.1$\pm$0.38 \\
\object{GRB 000926} & 2.037 & --20.82 & 0.85 & 9.55$\pm$4.91 \\
\object{GRB 010921} & 0.451 & --19.42 & 0.23 & 7.83$\pm$0.35\\
\noalign{\smallskip}
\hline
\end{tabular}
\caption{Absolute magnitudes, luminosities relative to an $L^*$
  galaxy, and specific SFRs of the GRB hosts. The specific SFRs are
  not corrected for internal extinction.}
\label{tab:sfrnorm}
\end{table}


\section{Comparison with field galaxies}
\label{comp}

\citet{lefloch03} found that GRB hosts are rather blue compared to other high
redshift galaxies, which could be interpreted as GRB hosts having higher SFRs,
or being less dusty than the average galaxy. We will now investigate whether
the GRB host SFRs are different from another sample of high redshift galaxies.
The SFRs for the field galaxies should be derived in the same manner as for
the GRB hosts.  We therefore need a large sample of high redshift galaxies
selected from optical methods and for which estimates of the redshifts exist.

\subsection{Comparison sample}
\label{sample_comp}
Magnitudes in the $U\!BV\!I\!J\!H\!K$ bands and photometric redshifts
of 1067 galaxies in the Hubble Deep Field North (HDFN) have been
estimated in
\citet{fernandez99}\footnote{http://bat.phys.unsw.edu.au/$\sim$fsoto/hdfcat.html}.
Data in the $U\!BV\!I$ bands were obtained with the WFPC2 using the
broad band filters F300W, F450W, F606W, and F814W, respectively.  The
$J\!H\!K$ data were from ground based photometry obtained with the 4m
telescope at the Kitt Peak Observatory.  Photometric redshifts are
uncertain within $\Delta$z=0.1 for the brightest galaxies with $I<25$
\citep{bolzo00}, estimated by comparing the photometric redshifts to
the spectroscopic redshifts of more than 100 galaxies in the
catalog. Since the galaxies have been observed in 7 bands and have
smaller photometric errors the photometric redshift accuracy is better
than for the GRB host sample. Additionally, a catalog containing 1611
galaxies, with optical data from WFPC2 and near-IR data from the
VLT/ISAAC of the HDF South (HDFS) was included
\citep{vanzella01}. Compared to the GRB hosts these galaxies, which we
will collectively refer to as HDF galaxies, have a wider span in
magnitudes and redshifts.

The flux densities in the various filters given in the catalogs were
converted into AB magnitudes which were used as input for HyperZ.
Conversion factors were calculated using information of the
throughputs of the WFPC2 filters for the optical data. Similarly, for
the near-IR data the throughputs of the Kitt Peak and VLT/ISAAC
filters were used, respectively.

Photometric redshifts, best fitting templates, extinctions, and absolute $B$
band magnitudes were estimated for all galaxies using the same cosmology as
for the GRB hosts.

In the redshift range corresponding to the redshift distribution of the GRB
hosts analysed here, $0.4<z<2.1$, the number of galaxies that have photometric
redshifts differing by more than 0.2 from the spectroscopic measure is 10\%.
This number represents a measure of the overall accuracy of the photometric
redshift estimations.

\begin{figure}
\resizebox{\hsize}{!}{\includegraphics[bb=35 0 580 405,clip]{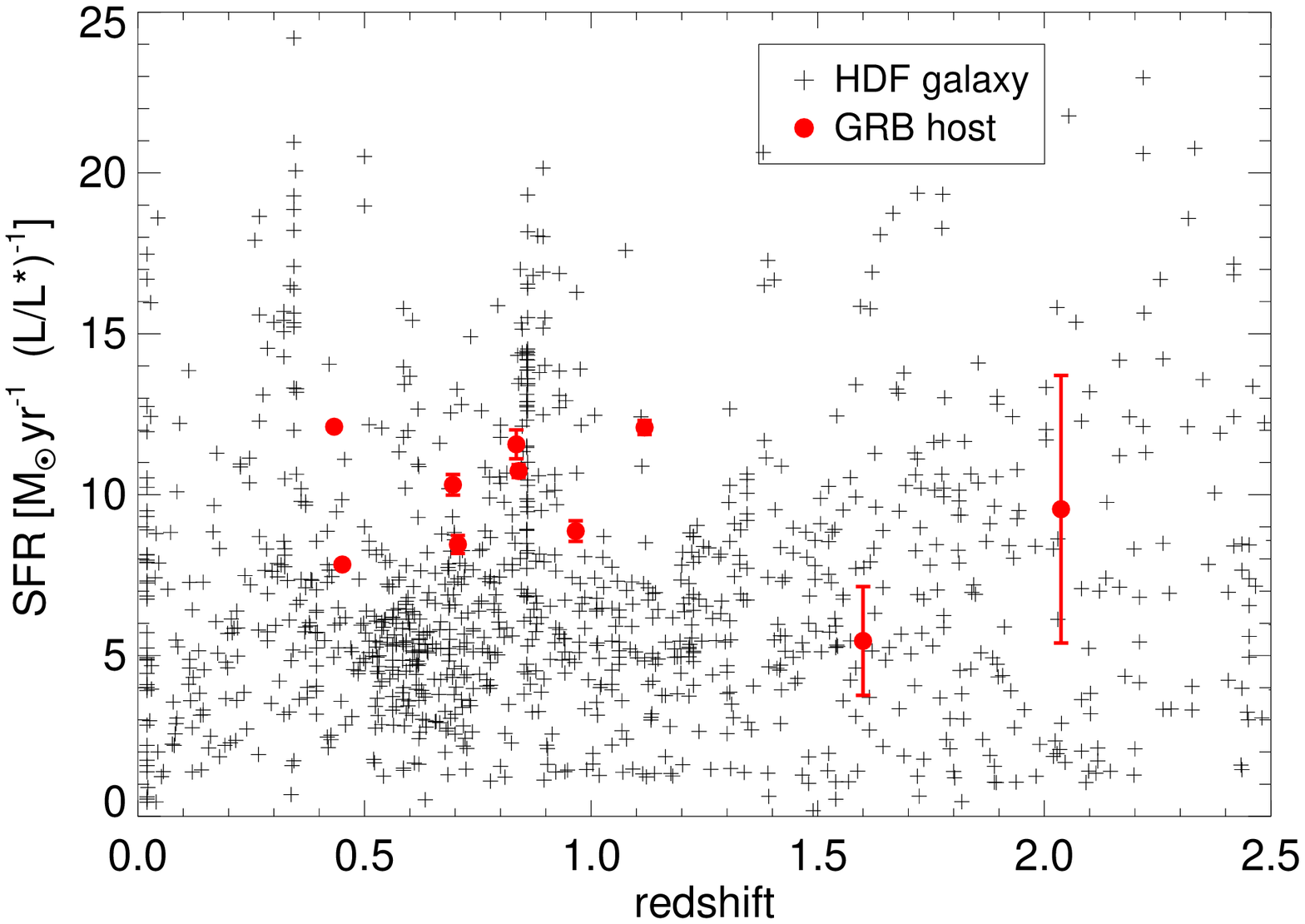}}
\resizebox{\hsize}{!}{\includegraphics[bb=35 0 570 405,clip]{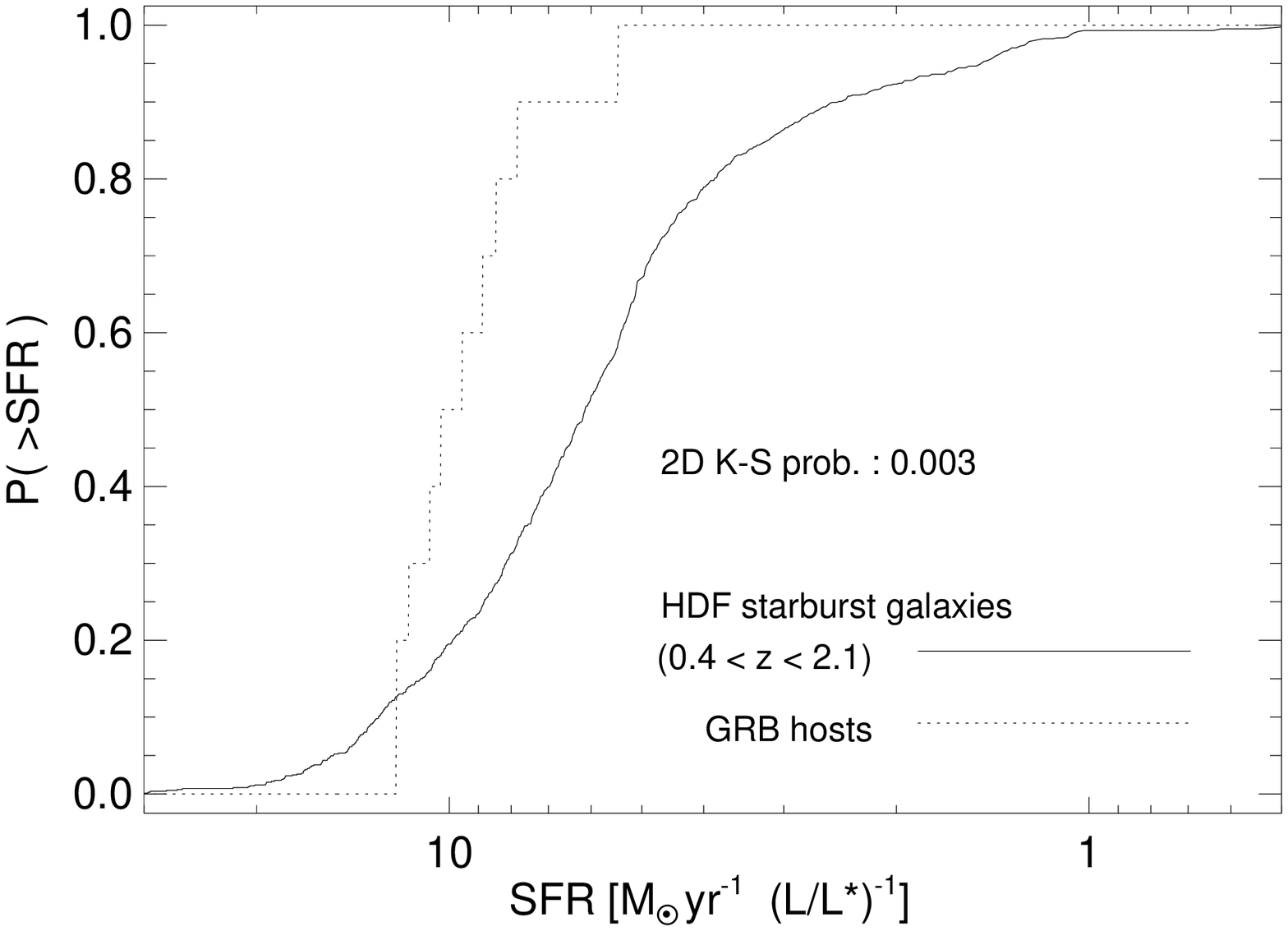}}
\caption{Upper panel: Specific SFRs for 1140 starburst galaxies in the HDF North
  and South. The galaxies have photometric redshifts estimated by HyperZ and
  are all classified as starburst galaxies. The SFRs of the 10 GRB hosts are
  shown as large circles. None of the SFRs are corrected for intrinsic
  extinction.  Lower panel: cumulative distribution of the specific SFR for
  851 starburst galaxies having $0.4<z<2.1$ (solid curve) and for 10 GRB hosts
  galaxies in the same redshift range (dotted curve).  67\% of the HDF
  galaxies classified as starburst galaxies (573 among 851) have specific SFRs
  of more than 5~M\subsun \peryr $(L/L^*)^{-1}$, which is the smallest
  specific SFR of all the GRB hosts.}
\label{fig:normsfrhdf}
\end{figure}

The SFRs of the 1067+1611 HDF galaxies were calculated in the same way
as described above for the GRB hosts. In total 1140 galaxies were
categorized as starburst galaxies, and for those the specific SFRs (in
M\subsun \peryr\ ($L/L^*)^{-1}$) were determined. As above, we assumed
$M^*~=~-21$ and did not correct for the effect of extinction. The
results are presented in Fig.~\ref{fig:normsfrhdf}. The top panel
shows the specific SFRs of 1140 HDF starburst galaxies as a function of
their estimated photometric redshifts. The specific SFRs (not
corrected for extinction) for the GRB hosts are shown as large
circles. Error bars are not included, but for each point the error is
$\sim$30\%, due to the intrinsic scatter of the SFR estimator.  All
specific SFRs of the HDF galaxies are in the range
0~--~20~M\subsun~\peryr\ $(L/L^*)^{-1}$. However relatively few
galaxies (20\%) have specific SFRs exceeding 10~M\subsun~\peryr\
$(L/L^*)^{-1}$, whereas this is the case for 50\% of the GRB hosts.

To perform a direct comparison with the sample of GRB hosts, 851 HDF
starburst galaxies with photometric redshifts in the range $0.4<z<2.1$
were selected.  The lower panel in Fig.~\ref{fig:normsfrhdf} shows the
cumulative distribution of the specific SFRs of these galaxies. On
average the specific SFRs for the GRB hosts are larger than for the
field galaxies. Out of the 851 HDF galaxies 573 galaxies have specific
SFRs above 5~M\subsun~\peryr\ $(L/L^*)^{-1}$, which is the lower range
of the SFRs of the GRB hosts. Taken at face value, this implies that
the population of {\em all} GRB hosts lie among the upper 67\% of
starburst galaxies, implying that GRB hosts have specific SFRs which
are larger than for ordinary field starburst galaxies at similar
redshifts.  Another explanation could be that the GRB hosts have less
extinction. We shall return to a discussion of this effect in
Sect.~\ref{sfr_comp} and Sect.~\ref{ext_comp}.

We performed a statistical test in order to determine whether the distribution
of specific SFRs vs. redshifts is different for the two samples. We applied a
{\em two-dimensional, two-sample Kolmogorov--Smirnov test} (K--S test)
\citep{peacock83} which uses two 2-dimensional samples and checks the
probability that one sample has the same parent distribution as the other. We
used the implementation of the test described in \citet{fasano87}, which uses
significantly fewer computations. According to \citet{peacock83} the test is
valid when both the sample sizes are greater than or equal to 10.
Applications of the tests, described in \citet{fasano87} and
\citet{peacock83}, have shown that there is no difference for uncorrelated
distributions, within statistical uncertainties, between the two tests.  When
the probability is $>0.2$ the value of the probability is not accurate, but
the hypothesis that the two distributions are not significantly different is
correct \citep{num}, and the derived probabilities can be considered as lower
limits.

Our qualitative finding that the distributions of specific SFRs vs.  redshifts
are different for the GRB hosts and HDF starburst galaxies is supported by the
2D K--S test, which gives a probability of 0.003 that the parent population is
the same for the two samples. Hereafter, when a two dimensional test is
performed, the first dimension corresponds to the redshift and the second to
the SFRs as in the upper panel in Fig.~\ref{fig:normsfrhdf}. Thus, we
calculate 2D probabilities for the distributions while showing the
corresponding 1D cumulative distribution, e.g. as in the lower panel in
Fig.~\ref{fig:normsfrhdf}.

\subsection{Population ages}
\label{pop_age}
\begin{figure}
\resizebox{\hsize}{!}{\includegraphics[bb=35 0 570 405,clip]{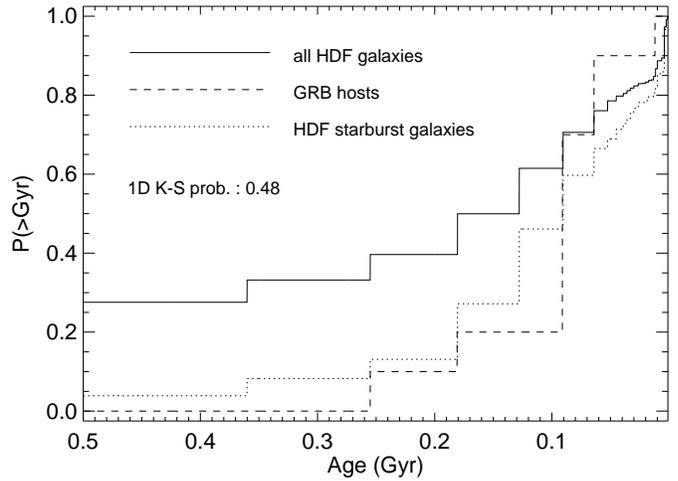}}
\caption{Age distribution of HDF galaxies and GRB host. The solid line shows
  the ages for all types of HDF galaxies in the redshift range $0.4<z<2.1$,
  while the dotted line shows the distribution of starburst galaxies only.
  The step-like appearance is due to the grid of ages of the available
  templates. A 1D K--S test give a probability of 48\% that the GRB host and
  HDF starburst galaxy distributions are the same.}
\label{fig:ages}
\end{figure}

In Sect.~\ref{sed} we found that GRB hosts are young starburst galaxies. We
therefore investigated whether they are younger on average than field
starburst galaxies. For comparison we used the ages for the 851 HDF starburst
galaxies. In Fig.~\ref{fig:ages} the distribution of ages of all HDF galaxies,
the GRB hosts and the HDF starburst galaxies are shown by the solid, dashed,
and dotted lines, respectively.

In addition to the two-dimensional, two sample test, we also use a
one-dimensional, two sample K--S test.  For the 1D two sample test, the
probabilities are reliable for sample sizes $N=N_1N_2/(N_1+N_2)>4$, where
$N_1$ and $N_2$ are the number of objects in sample 1 and 2, respectively.
This criterion is always satisfied for the tests performed in this paper.
Additionally, the probabilities are reliable in contrast with the 2D test.

Using a one-dimensional K-S test on the distribution of the 851 starburst
galaxies ages, we find a probability of 0.48 that the two distributions are
the same, whereas a 2D K-S test gives a probability of 0.18 that the
distributions of age vs.  redshifts are similar for the two populations.
Therefore, we have no clear indication that GRB hosts are on average younger
than field starburst galaxies.  We furthermore checked whether the redshift
distribution of the GRB hosts and the HDF starburst galaxies were the same
which is confirmed by a 1D K--S probability of 70\%.

Comparing the GRB host ages with those of all types of HDF galaxies at
redshifts $0.4<z<2.1$ we find a 1D K-S probability of 0.02 that the
distributions are similar. A comparison of the solid and the dashed lines in
Fig.~\ref{fig:ages} shows that GRB hosts are indeed younger than a sample of
all field galaxies.

\subsection{Very young galaxies}
\label{young_comp}
\begin{figure}
\resizebox{\hsize}{!}{\includegraphics[bb=35 0 570 405,clip]{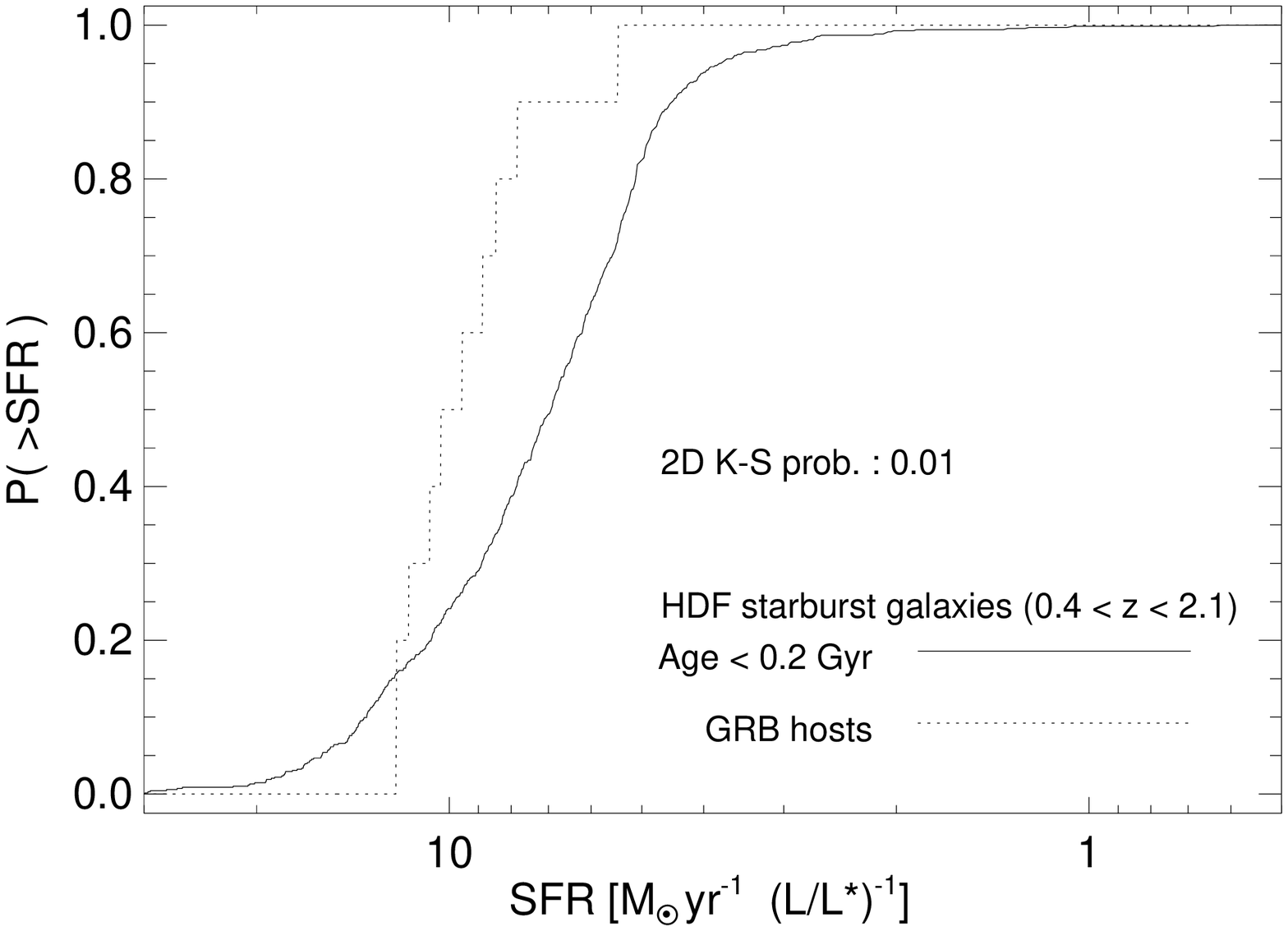}}
\resizebox{\hsize}{!}{\includegraphics[bb=35 0 570 405,clip]{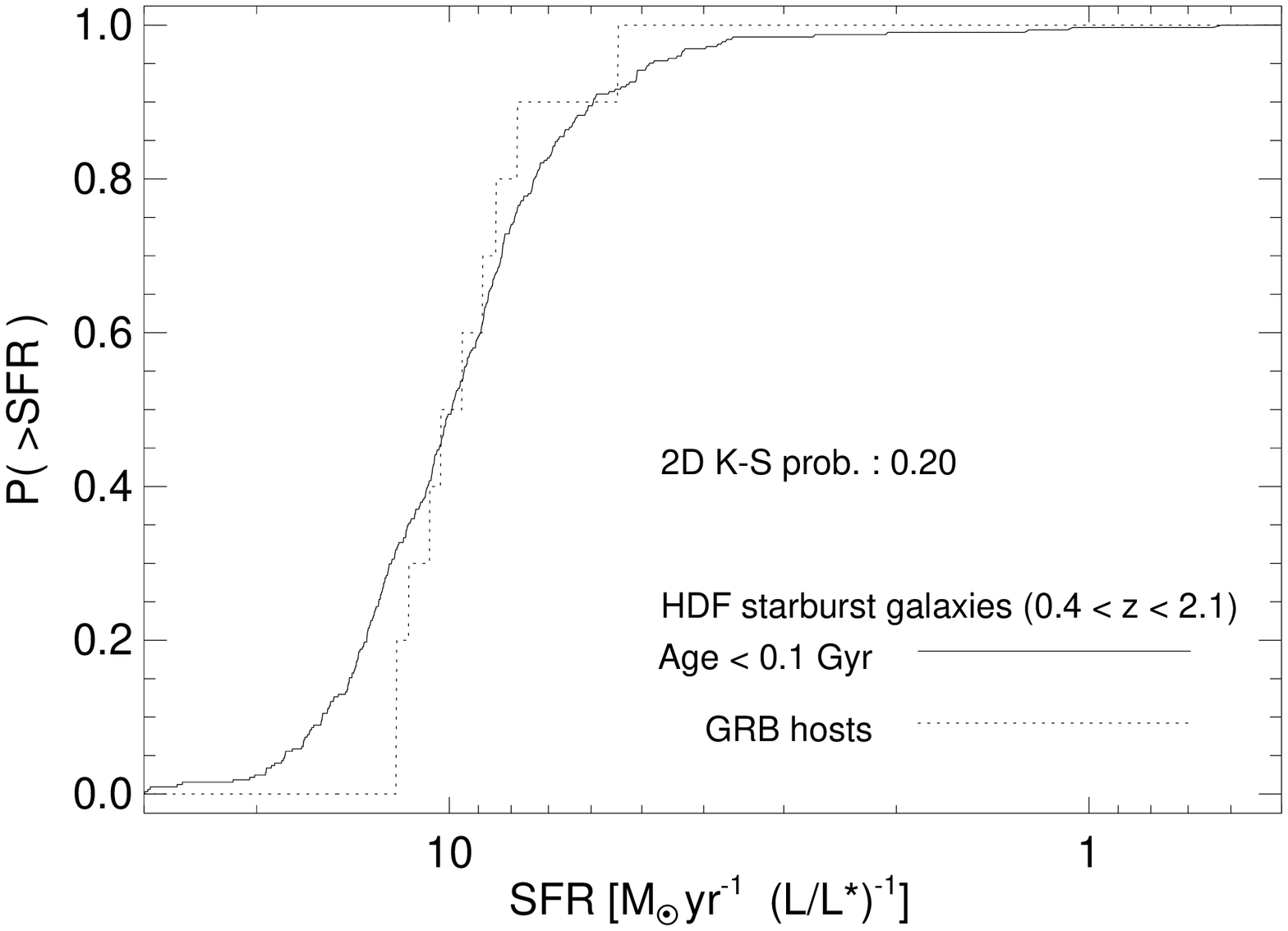}}
\caption{These plots are similar to the one in the lower panel in
  Fig.~\ref{fig:normsfrhdf}, but here only young starburst galaxies have been
  included. Upper panel: The solid line represents the specific SFR
  distribution of 689 HDF galaxies with ages less than 0.2 Gyr. 83\% of these
  HDF galaxies have specific SFRs larger than 5 M\subsun \peryr\ 
  $(L/L^*)^{-1}$. Applying the 2D K--S test gives a probability of 0.01 for
  the sample parent distribution. Lower panel: 327 HDF galaxies with ages less
  than 0.1 Gyr for which 94\% have specific SFRs larger than 5 M\subsun
  \peryr\ $(L/L^*)^{-1}$.  The two distributions are now rather similar, which
  is confirmed by the K--S probability of 0.20 for the same parent
  distribution.}
\label{fig:normyoung}
\end{figure}

The GRB hosts are classified by HyperZ as very young systems with ages smaller
than 0.2~Gyr. We therefore also selected a comparison sample consisting of
young (age $<0.2$~Gyr) HDF starburst galaxies. As expected, this changes the
picture since the selected HDF galaxies now have larger average SFRs as
indicated in the upper panel in Fig.~\ref{fig:normyoung}. The GRB host SFRs
now lie among the upper 83\% of the HDF galaxies. Applying the 2D K--S test
gives a probability of 0.01 for the same underlying distribution. If one
compares the GRB hosts to HDF galaxies classified as having ages smaller than
0.1 Gyr, the two populations become rather similar as seen in the lower panel
in Fig.~\ref{fig:normyoung}.  The 2D K--S test gives a probability of 0.20 for
the same parent distribution.

These simple tests show that the specific SFRs of GRB hosts are larger
on the average than a population of starburst galaxies at the same
redshifts. The HDF galaxies show a wider distribution than the GRB
hosts with more galaxies at the high and low end of the specific SFR
distribution, as indicated by the cumulative distribution shown in the
lower panel of Fig.~\ref{fig:normyoung}.

 Our finding that GRB hosts have larger specific star-formation
  rates than field galaxies gives observational support to the
  selection criteria of potential GRB hosts based on numerical
  simulations \citet{courty04}. Their selected GRB hosts do not have
  large absolute SFRs, but have high star formation efficiencies in
  agreement with our results.

\subsection{Comparisons with extinction corrected SFRs}
\label{sfr_comp}
\begin{figure}
  \resizebox{\hsize}{!}{\includegraphics[bb=35 0 570 405,clip]{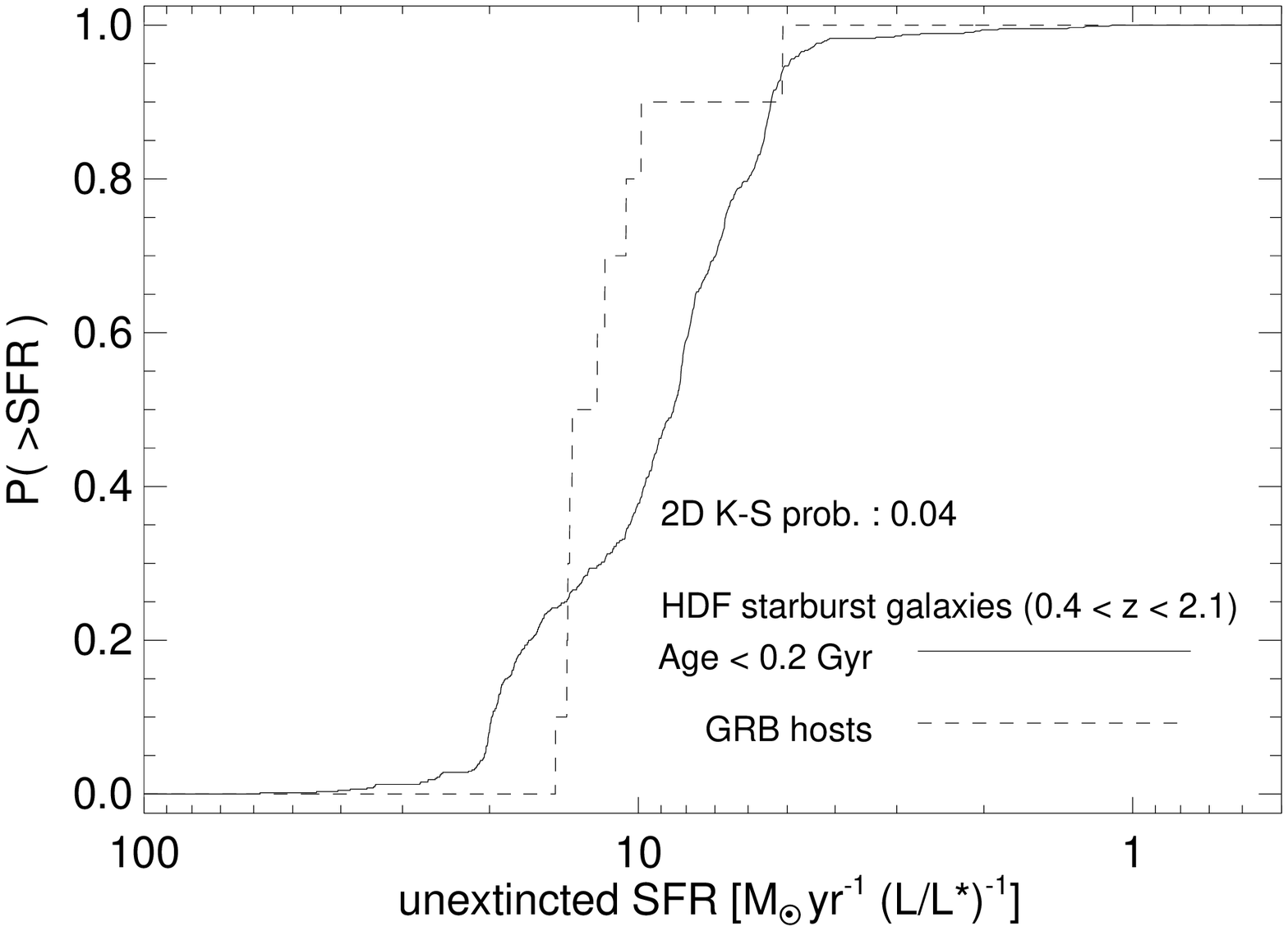}}
  \resizebox{\hsize}{!}{\includegraphics[bb=35 0 570 405,clip]{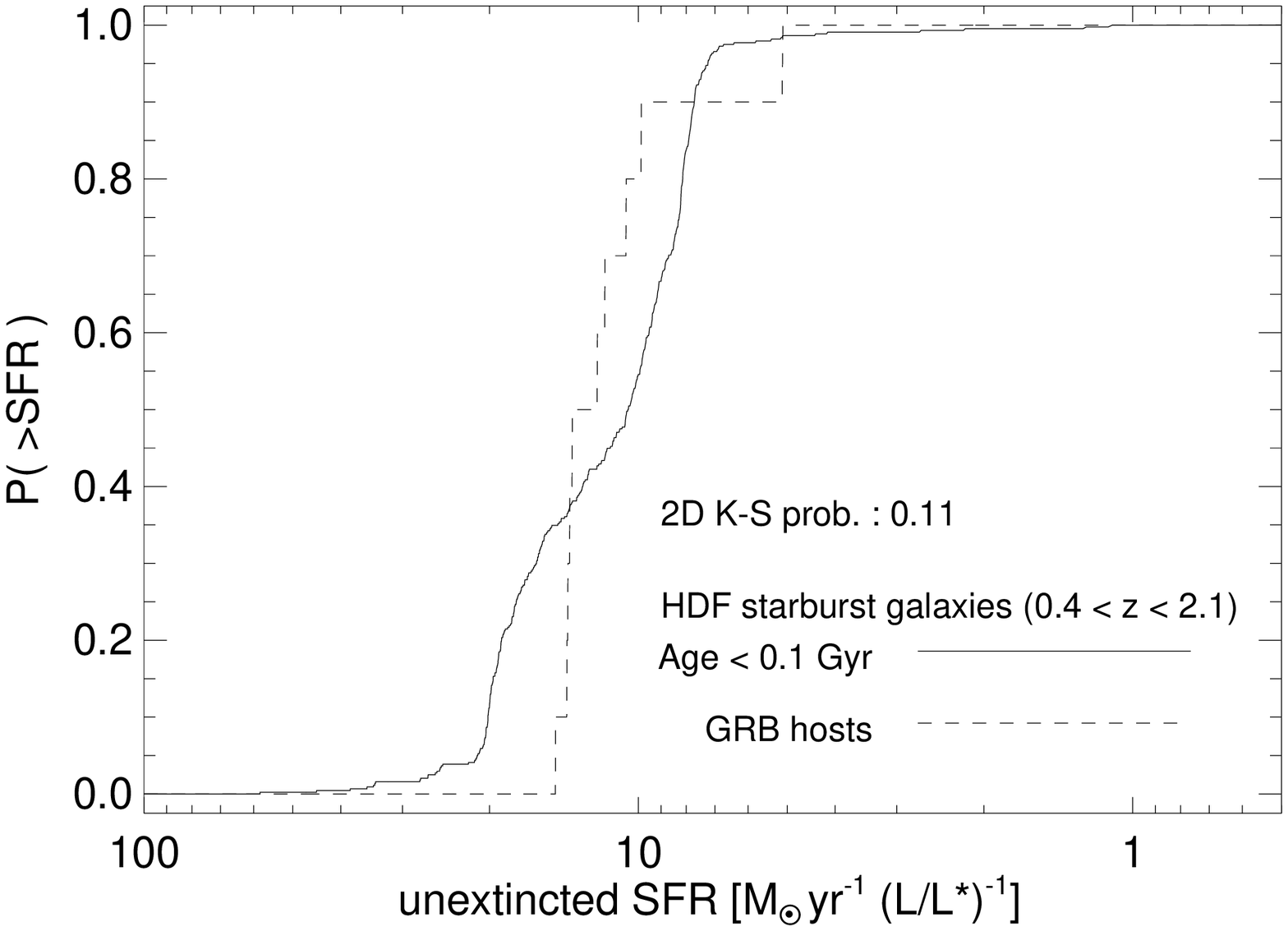}}
\caption{Cumulative distribution of the extinction corrected specific
  SFRs of the HDF starburst galaxies with $0.4<z<2.1$. The upper panel
  consists of HDF galaxies with ages smaller than 0.2 Gyr and the lower panel
  of galaxies with ages smaller than 0.1 Gyr. This again shows that GRB hosts
  are more likely to have similar specific SFRs as very young field starburst
  galaxies. The one dimensional K--S test gives the probabilities of 0.04 and
  0.11 for the same distribution in the two plots, respectively. }
\label{fig:sumext}
\end{figure}

The SFRs in Table \ref{tab:sfrnorm} were corrected for extinction and analysed
in the same manner as above. Similarly, the SFRs for the HDF galaxies were
corrected using the $A_V$s estimated by HyperZ. The absolute $B$ band
magnitudes of both galaxy samples were also corrected for extinction, using
$R_{4400}=5.06$ for a Calzetti extinction curve ($A_{4400}=R_{4400}$\ebv).
Fig.~\ref{fig:sumext} shows the cumulative distributions of these unextincted
SFRs, where the selection criteria are the same as in
Fig.~\ref{fig:normyoung}. The upper panel shows a comparison of GRB hosts with
young starburst (ages less than 0.2 Gyr) HDF galaxies. The 2D K--S test gives
a probability of 0.04 for the same parent distribution. The lower panel shows
the younger (age $<$0.1 Gyr) HDF galaxies, and the 2D K--S test gives a
probability of 0.11 for the same parent distribution.

All the calculated probabilities are given in Table~\ref{tab:ksprob}, using
either Miller \& Scalo or Salpeter based templates. Generally, the differences
between the two are small.  We can therefore conclude that GRB hosts are not
drawn at random from the average field starburst galaxy population and the GRB
hosts are most likely similar to HDF starburst galaxies with very young
($t<0.1$ Gyr) stellar populations. This conclusion is independent on the
assumed IMF and extinction correction.

\begin{table*}
\begin{center}
\begin{tabular}{lll|ll}
\\
\hline \hline
\noalign{\smallskip}
Age (Gyr) & M\&S IMF & Salpeter IMF  & M\&S IMF & Salpeter IMF \\
\noalign{\smallskip}
\hline
\noalign{\smallskip}
all     &  0.003 (851)& 0.006 (823) & 0.0001 (111) & 0.0005 (128)  \\
$<0.2$  &  0.01 (689) &  0.02 (692) & 0.0008 (85) & 0.007 (80)  \\
$<0.1$  &  0.20 (464) &  0.24 (444) & 0.003  (46) & 0.002 (42)\\
\noalign{\smallskip}
\hline
\noalign{\smallskip}
extinction corrected \\
\noalign{\smallskip}
\hline
\noalign{\smallskip}
all     &  0.006 (783) &  0.001 (794) & 0.05 (164) & 0.001 (218)\\
$<0.2$  &  0.04  (642) &  0.004 (676) & 0.06 (131) & 0.006 (164)\\
$<0.1$  &  0.11  (439) &  0.04 (440)  & 0.008 (84) & 0.004 (95)\\
\noalign{\smallskip}
\hline
\end{tabular}
\caption{2-dimensional K--S probabilities for the GRB hosts having the
  same parent distribution of specific SFRs vs. redshifts as the starburst HDF
  galaxies at similar redshifts ($0.4<z<2.1$) with the ages listed in column
  1. In brackets are given the number of galaxies which are used for
  comparison. Columns 2 and 3 impose no selection criteria on the brightness
  of the comparison galaxies, while column 4 and 5 include an additional
  criterion, that the HDF galaxies be as bright or brighter than absolute $B$
  band magnitude of the GRB hosts.}
\label{tab:ksprob}
\end{center}
\end{table*}

\subsection{Intrinsic extinction}
\label{ext_comp}

\begin{figure}
\resizebox{\hsize}{!}{\includegraphics[bb=35 0 570 405,clip]{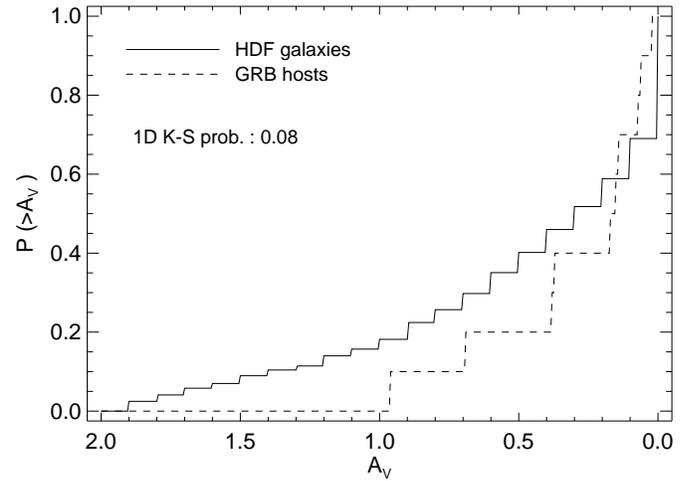}}
\caption{Cumulative distribution of the extinction values found for
  the GRB hosts (dotted line) and young ($<0.1$ Gyr) HDF field
  galaxies (solid line). In both samples a low extinction $A_V< 2$ is
  preferred by the best fits of the SEDs.  The two distributions
  appear different from each other, where the GRB hosts extinctions
  are on average smaller than for field galaxies,  but performing 1D
  K--S tests on the distributions gives inconclusive probabilities.}
\label{fig:ext}
\end{figure}

The probabilities for the extinction corrected SFR distributions are generally
smaller than for the same uncorrected SFR distributions which could indicate
that the extinctions found for the GRB hosts are different from those of young
field galaxies.  In Fig.~\ref{fig:ext} we show the cumulative distributions of
the extinction values found for the two samples. Extinction values for the
field galaxies are from the young (age $<0.1$ Gyr) HDF galaxies.  It seems
that the distributions are different in the sense that GRB hosts have on
average smaller extinctions compared to the young field galaxies. However,
performing one-dimensional K--S tests on the distributions of extinction
values yields probabilities of 0.29 and 0.08, for the same distribution, in
the case of the extinctions derived from a Miller \& Scalo IMF and a Salpeter
IMF, respectively. Therefore, the small GRB host sample does not allow us to
determine whether or not GRB hosts have smaller intrinsic extinctions than
young field starburst galaxies. Likewise, comparing GRB host extinctions with
those derived for all types of field galaxies does not allow us to determine
whether they have different extinction distributions.
  
\subsection{Additional tests}
\label{ad_comp}
As mentioned in Sect.~\ref{sed} the SED fit of the host of GRB~000418 could
indicate a somewhat larger extinction ($A_V=1.4$). We therefore investigated
the impact of changing this particular extinction value on the derived
probabilities for the extinction corrected specific SFRs. For a Salpeter IMF,
the derived 2D K--S probabilities for the same parent distribution are 0.03
and 0.04 for 0.2~Gyr and 0.1~Gyr old HDF starburst, respectively, while
Table~\ref{tab:ksprob} gives 0.004 and 0.04. We therefore conclude that the
uncertainty in the GRB~000418 host extinction has little impact on the
results.

We also investigated the possibility that other galaxy types might have
similar SFRs as the GRB hosts. Selecting young HDF galaxies classified as
other spectral types than starbursts, i.e. irregular galaxies or spirals with
$t<0.1$ Gyr, gives a 2D K--S probability of 0.002 for the same parent
distribution. We note that for the theoretical templates the differences
between the templates at such young ages are small. However, comparing the
specific extinction corrected SFRs with those of ellipticals, without any age
constraint, the probability is also small (2$\times10^{-5}$) which is not
surprising as old elliptical galaxies do not have any star formation.

Because of possible differences between bright and faint HDF galaxies, an
additional brightness requirement on the HDF galaxies was therefore imposed.
We selected HDF galaxies with absolute $B$ band magnitudes as bright or
brighter than GRB hosts ($M_B < -18.08$) and the resulting K--S probabilities
for the various tests are listed in column 4 and 5 in Table~\ref{tab:ksprob}.
The probabilities now appear to be much smaller than without the brightness
selection (apart from the M\&S IMF based extinction corrected tests). This is
not due to the smaller number of comparison galaxies, but arises because the
distribution of specific SFRs for the GRB hosts is narrower than for the HDF
galaxies as also seen in Figs.~\ref{fig:normyoung} and \ref{fig:sumext}.
However, we consider that an increase in the GRB host sample is necessary
before this difference in the distributions can be explained.

\section{Discussion and conclusion}
\label{conc}

We have constructed a flux-limited sample ($R < 25.3$) of all GRB host
galaxies with known redshifts.  The sample consists of 10 galaxies with
broad-band magnitudes in more than 5 filters obtained from the literature.
The GRBs which occurred in these galaxies comprise a collection of a dark
burst (GRB~000210), a dim one (GRB~000418), a bright one (GRB~990712), and a
very bright one (GRB~990123).

Comparing the SEDs of the GRB hosts with template spectra we find that they
are young starburst galaxies with moderate to low extinctions ($A_V<1$).
Photometric redshifts are found to be accurate, with a standard deviation of
$\pm0.21$ from the spectroscopic ones.  Accurate photometric redshifts are
obtained provided there is sufficient optical-IR coverage and the magnitudes
are accurate to the 10--20\% level. Through the analysis of the SEDs of GRB
hosts we have found that it is important to include near-IR magnitudes when
estimating the extinction, since the effect of extinction is largest in the UV
region, and a better sampling of the broad band SEDs of the hosts gives a more
secure estimate of the extinction.

By comparing extinctions derived from the SEDs with those of galaxies in the
HDF North and South, we found that the intrinsic extinctions of the GRB hosts
are small and not significantly different on average from those of either
young field starburst galaxies or field galaxies in general.

The coincidence between small values of $A_V$ from the host SED and that of
the afterglow suggests that we mostly see effects of the global extinction in
the afterglow light curves. Moreover, small extinction values do not exclude
the possibility that the GRBs themselves are located in more dusty and higher
density environments, such as embedded in molecular clouds which has been
suggested through analyses of X-ray afterglows.  This was discussed for the
case of the GRB~000210 host galaxy, where a large $N_H$ was inferred
\citep{piro02}, while the galaxy itself shows a small global extinction (see
Table \ref{tab:hypsed}, and \citet{goro02}).  However, if the regions of star
formation where GRBs occur are small and not numerous, this will not have a
large effect on the overall SEDs of the hosts.

For all host galaxies the inferred ages are less than 0.2 Gyr while most
galaxies have even younger populations, $t<0.1$ Gyr.  A comparison of GRB host
ages with those of HDF galaxies showed that GRB hosts are not significantly
younger than starburst field galaxies at similar redshifts, but are
  clearly younger than a sample of all types of field galaxies.

  A good sampling of the redshifted Balmer jump/4000 {\AA} break gives a well
  determined age for the dominant population of stars in the galaxy. With
  multiband photometry this jump is sampled well for all galaxies in the
  redshift range involved in this study which indicates that the ages of the
  dominant population of stars are well constrained. Through fits to templates
  of various metallicities we find that different template metallicities give
  consistent estimates for the derived best fit ages.  Specifically, for 2
  hosts the ages varied by a factor of $\lesssim2$, while for the 8 remaining
  hosts, the ages were consistent.

The SFRs of the hosts were calculated from their rest frame 2800~{\AA} flux
and was found to vary by more than an order of magnitude from host to host.
Specific SFRs, obtained by normalising the SFRs with respect to the
luminosities of the hosts are more clustered around the mean value ranging
from 5 to 12 M\subsun \peryr\, $(L/L^*)^{-1}$ independently of the redshift.
Comparing these with specific SFRs of high redshift galaxies in the HDF we
found that the specific SFRs for GRB hosts lie among the upper 66\% of the
specific SFRs for the field galaxies in the same redshift range ($0.4<z<2.1$).
We performed several two-dimensional K--S tests to quantify the comparisons of
GRB hosts and subsets of the HDF field galaxy sample. We found that GRB hosts
most likely have specific SFRs similar to very young field galaxies with ages
less than 0.1 Gyr. Taking extinction effects into account does not change this
result. The inferred young ages of the dominant stellar populations of the GRB
hosts are in agreement with the idea that GRBs are associated with core
collapse SNe \citep{woosley93,gal98,hjorth03}.

We have found that GRB hosts are not younger than field starburst galaxies but
have similar specific SFRs as the youngest starburst galaxies showing that GRB
hosts belong to a group of very young, actively star forming galaxies.

The ages are inferred from the size of the Balmer jump/4000 {\AA} break,
but also from the slope of the spectrum, while the specific SFRs measure the
flux ratio between 4400{\AA} and 2800 {\AA} in the rest frame. All SED fits
were done with templates of solar metallicity, which is likely a
simplification.

Some GRB hosts are found to be Lyman$\alpha$ emitters indicating that these
galaxies contain only little dust or have low metallicities \citep{fynbo03}.
Low dust content and low metallicity of the environment is also indicated by
spectroscopic observations of the optical afterglow of \object{GRB~020124}
\citep{hjorth03b}. Low internal metallicity for GRB hosts would imply that we
observe bluer colours relative to the HDF galaxies which give rise to larger
specific SFRs and furthermore, the SED fits would result in a younger age
assuming solar metallicity.  The conclusions that GRB hosts have similar ages
as field starburst galaxy and yet appear to have larger specific SFRs may
therefore be consistent.

A larger sample of GRB hosts can be constructed by obtaining multiband
observations of hosts of bursts which have occurred within the past
two years. This can be used to analyse SEDs and infer SFRs from
individual galaxies along the lines presented in this paper. Moreover,
with future space based missions, such as Swift, a sample of uniformly
selected GRBs with sufficiently brights host galaxies is within
reach. This would allow a detailed quantitative comparison with the
properties of specific subsamples of optically selected field galaxies
at high redshift.

\begin{acknowledgements} 
  L.~Christensen acknowledges support by the German Verbundforschung
  associated with the ULTROS project, grant no. 05AE2BAA/4.  We are grateful
  to Johan Fynbo for providing a preliminary GRB 000926 near-infrared host
  magnitude prior to publication. The authors acknowledge benefits from
  collaboration with the EU FP5 Research Training Network ``Gamma Ray Bursts:
  An Enigma and a Tool''. This work was supported by the Danish Natural
  Science Research Council (SNF). We finally thank the anonymous referee for
  comments and suggestions which helped to improve the paper.
\end{acknowledgements}
\bibliography{P_0361}

\begin{thebibliography}{73}
\expandafter\ifx\csname natexlab\endcsname\relax\def\natexlab#1{#1}\fi

\bibitem[{Andersen {et~al.}(2000)Andersen, Hjorth, Pedersen, Jensen, Hunt,
  Gorosabel, M{\o}ller, Fynbo, Kippen, Thomsen, Olsen, Christensen,
  Vestergaard, Masetti, Palazzi, Hurley, Cline, Kaper, \& Jaunsen}]{andersen00}
Andersen, M.~I., Hjorth, J., Pedersen, H., {et~al.} 2000, A\&A, 364, L54

\bibitem[{{Berger} {et~al.}(2003){Berger}, {Cowie}, {Kulkarni}, {Frail},
  {Aussel}, \& {Barger}}]{berger02}
{Berger}, E., {Cowie}, L.~L., {Kulkarni}, S.~R., {et~al.} 2003, \apj, 588, 99

\bibitem[{{Berger} {et~al.}(2002){Berger}, {Kulkarni}, {Bloom}, {Price}, {Fox},
  {Frail}, {Axelrod}, {Chevalier}, {Colbert}, {Costa}, {Djorgovski},
  {Frontera}, {Galama}, {Halpern}, {Harrison}, {Holtzman}, {Hurley}, {Kimble},
  {McCarthy}, {Piro}, {Reichart}, {Ricker}, {Sari}, {Schmidt}, {Wheeler},
  {Vanderppek}, \& {Yost}}]{berger02b}
{Berger}, E., {Kulkarni}, S.~R., {Bloom}, J.~S., {et~al.} 2002, \apj, 581, 981

\bibitem[{Berger {et~al.}(2001)Berger, Kulkarni, \& Frail}]{berger01}
Berger, E., Kulkarni, S.~R., \& Frail, D.~A. 2001, ApJ, 560, 652

\bibitem[{{Bloom} {et~al.}(2003){Bloom}, {Berger}, {Kulkarni}, {Djorgovski}, \&
  {Frail}}]{bloom03}
{Bloom}, J.~S., {Berger}, E., {Kulkarni}, S.~R., {Djorgovski}, S.~G., \&
  {Frail}, D.~A. 2003, \aj, 125, 999

\bibitem[{{Bloom} {et~al.}(2001){Bloom}, {Djorgovski}, \& {Kulkarni}}]{bloom01}
{Bloom}, J.~S., {Djorgovski}, S.~G., \& {Kulkarni}, S.~R. 2001, ApJ, 554, 678

\bibitem[{{Bloom} {et~al.}(1998){Bloom}, {Djorgovski}, {Kulkarni}, \&
  {Frail}}]{bloom98b}
{Bloom}, J.~S., {Djorgovski}, S.~G., {Kulkarni}, S.~R., \& {Frail}, D.~A. 1998,
  \apjl, 507, L25

\bibitem[{{Bloom} {et~al.}(2002){Bloom}, {Kulkarni}, \& {Djorgovski}}]{bloom00}
{Bloom}, J.~S., {Kulkarni}, S.~R., \& {Djorgovski}, S.~G. 2002, AJ, 123, 1111

\bibitem[{{Bloom} {et~al.}(1999){Bloom}, {Kulkarni}, {Djorgovski},
  {Eichelberger}, {Cote}, {Blakeslee}, {Odewahn}, {Harrison}, {Frail},
  {Filippenko}, {Leonard}, {Riess}, {Spinrad}, {Stern}, {Bunker}, {Dey},
  {Grossan}, {Perlmutter}, {Knop}, {Hook}, \& {Feroci}}]{bloom99}
{Bloom}, J.~S., {Kulkarni}, S.~R., {Djorgovski}, S.~G., {et~al.} 1999, Nature,
  401, 453

\bibitem[{{Bolzonella} {et~al.}(2000){Bolzonella}, {Miralles}, \& {Pell{\'
  o}}}]{bolzo00}
{Bolzonella}, M., {Miralles}, J.-M., \& {Pell{\' o}}, R. 2000, A\&A, 363, 476

\bibitem[{{Bruzual} \& {Charlot}(1993)}]{bru93}
{Bruzual}, A.~G. \& {Charlot}, S. 1993, ApJ, 405, 538

\bibitem[{{Bruzual} \& {Charlot}(2003)}]{bruzual03}
{Bruzual}, A.~G. \& {Charlot}, S. 2003, \mnras, 344, 1000

\bibitem[{{Calzetti} {et~al.}(2000){Calzetti}, {Armus}, {Bohlin}, {Kinney},
  {Koornneef}, \& {Storchi-Bergmann}}]{calz00}
{Calzetti}, D., {Armus}, L., {Bohlin}, R.~C., {et~al.} 2000, ApJ, 533, 682

\bibitem[{{Castro} {et~al.}(2003){Castro}, {Galama}, {Harrison}, {Holtzman},
  {Bloom}, {Djorgovski}, \& {Kulkarni}}]{castro03}
{Castro}, S., {Galama}, T.~J., {Harrison}, F.~A., {et~al.} 2003, \apj, 586, 128

\bibitem[{{Castro-Tirado} \& {Gorosabel}(1999)}]{ctir99}
{Castro-Tirado}, A.~J. \& {Gorosabel}, J. 1999, A\&AS, 138, 449

\bibitem[{{Castro-Tirado} {et~al.}(2001){Castro-Tirado}, {Sokolov},
  {Gorosabel}, {Castro Cer{\' o}n}, {Greiner}, {Wijers}, {Jensen}, {Hjorth},
  {Toft}, {Pedersen}, {Palazzi}, {Pian}, {Masetti}, {Sagar}, {Mohan}, {Pandey},
  {Pandey}, {Dodonov}, {Fatkhullin}, {Afanasiev}, {Komarova}, {Moiseev},
  {Hudec}, {Simon}, {Vreeswijk}, {Rol}, {Klose}, {Stecklum}, {Zapatero-Osorio},
  {Caon}, {Blake}, {Wall}, {Heinlein}, {Henden}, {Benetti}, {Magazz{\` u}},
  {Ghinassi}, {Tommasi}, {Bremer}, {Kouveliotou}, {Guziy}, {Shlyapnikov},
  {Hopp}, {Feulner}, {Dreizler}, {Hartmann}, {Boehnhardt}, {Paredes}, {Mart{\'
  i}}, {Xanthopoulos}, {Kristen}, {Smoker}, \& {Hurley}}]{cas01}
{Castro-Tirado}, A.~J., {Sokolov}, V.~V., {Gorosabel}, J., {et~al.} 2001, A\&A,
  370, 398

\bibitem[{Castro-Tirado {et~al.}(1999)Castro-Tirado, Zapatero-Osorio, Caon,
  Cairós, Hjorth, Pedersen, Andersen, Gorosabel, Bartolini, Guarnieri,
  Piccioni, Frontera, Masetti, Palazzi, Pian, Greiner, Hudec, Sagar, Pandey,
  Mohan, Yadav, Nilakshi, Björnsson, Jakobsson, Burud, Courbin, Valentini,
  Piersimoni, Aceituno, Montoya, Pedraz, Gredel, Claver, Rector, Rhoads,
  Walter, Ott, Hippelein, Sánchez-Béjar, Gutiérrez, Oscoz, Zhu, Chen, Zhang,
  Wei, Zhou, Guziy, Shlyapnikov, Heise, Costa, Feroci, , \& Piro}]{castro99}
Castro-Tirado, A.~J., Zapatero-Osorio, M.~R., Caon, N., {et~al.} 1999, Science,
  283, 2069

\bibitem[{{Chary} {et~al.}(2002){Chary}, {Becklin}, \& {Armus}}]{chary01}
{Chary}, R., {Becklin}, E.~E., \& {Armus}, L. 2002, ApJ, 566, 229

\bibitem[{{Christensen} {et~al.}(2004){Christensen}, {Hjorth}, {Gorosabel},
  {Vreeswijk}, {Fruchter}, {Sahu}, \& {Petro}}]{lise03}
{Christensen}, L., {Hjorth}, J., {Gorosabel}, J., {et~al.} 2004, A\&A, 413, 121

\bibitem[{{Courty} {et~al.}(2004){Courty}, {Bj\"ornsson}, \&
  {Gudmundsson}}]{courty04}
{Courty}, S., {Bj\"ornsson}, G., \& {Gudmundsson}, E.~H. 2004, MNRAS, in press

\bibitem[{{de Grijs} {et~al.}(2003){de Grijs}, {Fritze-v.~Alvensleben},
  {Anders}, {Gallagher}, {Bastian}, {Taylor}, \& {Windhorst}}]{degrijs03}
{de Grijs}, R., {Fritze-v.~Alvensleben}, U., {Anders}, P., {et~al.} 2003,
  \mnras, 342, 259

\bibitem[{{Djorgovski} {et~al.}(2003){Djorgovski}, {Bloom}, \&
  {Kulkarni}}]{djorgov03}
{Djorgovski}, S.~G., {Bloom}, J.~S., \& {Kulkarni}, S.~R. 2003, \apjl, 591, L13

\bibitem[{{Djorgovski} {et~al.}(2001{\natexlab{a}}){Djorgovski}, {Frail},
  {Kulkarni}, {Bloom}, {Odewahn}, \& {Diercks}}]{djor01b}
{Djorgovski}, S.~G., {Frail}, D.~A., {Kulkarni}, S.~R., {et~al.}
  2001{\natexlab{a}}, ApJ, 562, 654

\bibitem[{{Djorgovski} {et~al.}(2001{\natexlab{b}}){Djorgovski}, {Kulkarni},
  {Bloom}, {Frail}, {Harrison}, {Galama}, {Reichart}, {Castro}, {Fox}, {Sari},
  {Berger}, {Price}, {Yost}, {Goodrich}, \& {Chaffee}}]{djor01c}
{Djorgovski}, S.~G., {Kulkarni}, S.~R., {Bloom}, J.~S., {et~al.}
  2001{\natexlab{b}}, in Gamma-ray Bursts in the Afterglow Era, ed. E.~{Costa},
  F.~{Frontera}, \& J.~{Hjorth} (Berlin Heidelberg: Springer), 218

\bibitem[{{Djorgovski} {et~al.}(1998){Djorgovski}, {Kulkarni}, {Bloom},
  {Goodrich}, {Frail}, {Piro}, \& {Palazzi}}]{djor98}
{Djorgovski}, S.~G., {Kulkarni}, S.~R., {Bloom}, J.~S., {et~al.} 1998, \apjl,
  508, L17

\bibitem[{{Fasano} \& {Franceschini}(1987)}]{fasano87}
{Fasano}, G. \& {Franceschini}, A. 1987, \mnras, 225, 155

\bibitem[{{Fern{\' a}ndez-Soto} {et~al.}(2001){Fern{\' a}ndez-Soto},
  {Lanzetta}, {Chen}, {Pascarelle}, \& {Yahata}}]{fernandez01}
{Fern{\' a}ndez-Soto}, A., {Lanzetta}, K.~M., {Chen}, H., {Pascarelle}, S.~M.,
  \& {Yahata}, N. 2001, \apjs, 135, 41

\bibitem[{{Fern{\' a}ndez-Soto} {et~al.}(1999){Fern{\' a}ndez-Soto},
  {Lanzetta}, \& {Yahil}}]{fernandez99}
{Fern{\' a}ndez-Soto}, A., {Lanzetta}, K.~M., \& {Yahil}, A. 1999, ApJ, 513, 34

\bibitem[{{Fitzpatrick}(1986)}]{fitz86}
{Fitzpatrick}, E.~L. 1986, AJ, 92, 1068

\bibitem[{Fruchter {et~al.}(1999{\natexlab{a}})Fruchter, Pian, Thorsett,
  Bergeron, Gonz$\acute{a}$lez, Metzger, Goudfrooij, Sahu, Ferguson, Livio,
  Mutchler, Petro, Frontera, Galama, Groot, Hook, Kouveliotou, Macchetto, van
  Paradijs, Palazzi, Pedersen, Sparks, \& Tavani}]{fruchter99}
Fruchter, A.~S., Pian, E., Thorsett, S.~E., {et~al.} 1999{\natexlab{a}}, ApJ,
  516, 683

\bibitem[{Fruchter {et~al.}(1999{\natexlab{b}})Fruchter, Thorsett, Metzger,
  Sahu, Petro, Livio, Ferguson, Pian, Hogg, Galama, Gull, Kouveliotou,
  Macchetto, van Paradijs, Pedersen, \& Smette}]{fruc99b}
Fruchter, A.~S., Thorsett, S.~E., Metzger, M.~R., {et~al.} 1999{\natexlab{b}},
  \apjl, 519, L13

\bibitem[{{Fynbo} {et~al.}(2003){Fynbo}, {Jakobsson}, {M{\o}ller}, {Hjorth},
  {Thomsen}, {Andersen}, {Fruchter}, {Gorosabel}, {Holland}, {Ledoux},
  {Pedersen}, {Rhoads}, {Weidinger}, \& {Wijers}}]{fynbo03}
{Fynbo}, J.~P.~U., {Jakobsson}, P., {M{\o}ller}, P., {et~al.} 2003, \aap, 406,
  L63

\bibitem[{{Fynbo} {et~al.}(2000){Fynbo}, {Holland}, {Andersen}, {Thomsen},
  {Hjorth}, {Bj{\" o}rnsson}, {Jaunsen}, {Natarajan}, \& {Tanvir}}]{fynbo00}
{Fynbo}, J.~U., {Holland}, S., {Andersen}, M.~I., {et~al.} 2000, \apjl, 542,
  L89

\bibitem[{{Galama} {et~al.}(2000){Galama}, {Tanvir}, {Vreeswijk}, {Wijers},
  {Groot}, {Rol}, {van Paradijs}, {Kouveliotou}, {Fruchter}, {Masetti},
  {Pedersen}, {Margon}, {Deutsch}, {Metzger}, {Armus}, {Klose}, \&
  {Stecklum}}]{gal00b}
{Galama}, T.~J., {Tanvir}, N., {Vreeswijk}, P.~M., {et~al.} 2000, ApJ, 536, 185

\bibitem[{{Galama} {et~al.}(1998){Galama}, {Vreeswijk}, {van Paradijs},
  {Kouveliotou}, {Augusteijn}, {Bohnhardt}, {Brewer}, {Doublier}, {Gonzalez},
  {Leibundgut}, {Lidman}, {Hainaut}, {Patat}, {Heise}, {in 't Zand}, {Hurley},
  {Groot}, {Strom}, {Mazzali}, {Iwamoto}, {Nomoto}, {Umeda}, {Nakamura},
  {Young}, {Suzuki}, {Shigeyama}, {Koshut}, {Kippen}, {Robinson}, {de Wildt},
  {Wijers}, {Tanvir}, {Greiner}, {Pian}, {Palazzi}, {Frontera}, {Masetti},
  {Nicastro}, {Feroci}, {Costa}, {Piro}, {Peterson}, {Tinney}, {Boyle},
  {Cannon}, {Stathakis}, {Sadler}, {Begam}, \& {Ianna}}]{gal98}
{Galama}, T.~J., {Vreeswijk}, P.~M., {van Paradijs}, J., {et~al.} 1998, Nature,
  395, 670

\bibitem[{{Garnavich} {et~al.}(2003){Garnavich}, {Stanek}, {Wyrzykowski},
  {Infante}, {Bendek}, {Bersier}, {Holland}, {Jha}, {Matheson}, {Kirshner},
  {Krisciunas}, {Phillips}, \& {Carlberg}}]{garnavich03}
{Garnavich}, P.~M., {Stanek}, K.~Z., {Wyrzykowski}, L., {et~al.} 2003, \apj,
  582, 924

\bibitem[{{Gorosabel} {et~al.}(2003{\natexlab{a}}){Gorosabel}, {Christensen},
  {Hjorth}, {Fynbo}, {Pedersen}, {Jensen}, {Andersen}, {Lund}, {Jaunsen},
  {Castro Cer{\' o}n}, {Castro-Tirado}, {Fruchter}, {Greiner}, {Pian},
  {Vreeswijk}, {Burud}, {Frontera}, {Kaper}, {Klose}, {Kouveliotou}, {Masetti},
  {Palazzi}, {Rhoads}, {Rol}, {Salamanca}, {Tanvir}, {Wijers}, \& {van den
  Heuvel}}]{goro02}
{Gorosabel}, J., {Christensen}, L., {Hjorth}, J., {et~al.} 2003{\natexlab{a}},
  \aap, 400, 127

\bibitem[{{Gorosabel} {et~al.}(2003{\natexlab{b}}){Gorosabel}, {Klose},
  {Christensen}, {Fynbo}, {Hjorth}, {Greiner}, {Tanvir}, {Jensen}, {Pedersen},
  {Holland}, {Lund}, {Jaunsen}, {Castro Cer{\' o}n}, {Castro-Tirado},
  {Fruchter}, {Pian}, {Vreeswijk}, {Burud}, {Frontera}, {Kaper}, {Kouveliotou},
  {Masetti}, {Palazzi}, {Rhoads}, {Rol}, {Salamanca}, {Wijers}, \& {van den
  Heuvel}}]{goro03}
{Gorosabel}, J., {Klose}, S., {Christensen}, L., {et~al.} 2003{\natexlab{b}},
  \aap, 409, 123

\bibitem[{{Greiner} {et~al.}(2003){Greiner}, {Klose}, {Salvato}, {Zeh},
  {Schwarz}, {Hartmann}, {Masetti}, {Stecklum}, {Lamer}, {Lodieu}, {Scholz},
  {Sterken}, {Gorosabel}, {Burud}, {Rhoads}, {Mitrofanov}, {Litvak}, {Sanin},
  {Grinkov}, {Andersen}, {Castro Cer{\' o}n}, {Castro-Tirado}, {Fruchter},
  {Fynbo}, {Hjorth}, {Kaper}, {Kouveliotou}, {Palazzi}, {Pian}, {Rol},
  {Tanvir}, {Vreeswijk}, {Wijers}, \& {van den Heuvel}}]{greiner03}
{Greiner}, J., {Klose}, S., {Salvato}, M., {et~al.} 2003, \apj, 599, 1223

\bibitem[{{Hjorth} {et~al.}(2000{\natexlab{a}}){Hjorth}, {Holland}, {Courbin},
  {Dar}, {Olsen}, \& {Scodeggio}}]{hjorth00b}
{Hjorth}, J., {Holland}, S., {Courbin}, F., {et~al.} 2000{\natexlab{a}}, \apjl,
  539, L75

\bibitem[{{Hjorth} {et~al.}(2000{\natexlab{b}}){Hjorth}, {Holland}, {Courbin},
  {Dar}, {Olsen}, \& {Scodeggio}}]{hjorth00}
{Hjorth}, J., {Holland}, S., {Courbin}, F., {et~al.} 2000{\natexlab{b}}, \apjl, 534, L147

\bibitem[{{Hjorth} {et~al.}(2003{\natexlab{a}}){Hjorth}, {M{\o}ller},
  {Gorosabel}, {Fynbo}, {Toft}, {Jaunsen}, {Kaas}, {Pursimo}, {Torii}, {Kato},
  {Yamaoka}, {Yoshida}, {Thomsen}, {Andersen}, {Burud}, {Cer{\' o}n},
  {Castro-Tirado}, {Fruchter}, {Kaper}, {Kouveliotou}, {Masetti}, {Palazzi},
  {Pedersen}, {Pian}, {Rhoads}, {Rol}, {Tanvir}, {Vreeswijk}, {Wijers}, \& {van
  den Heuvel}}]{hjorth03b}
{Hjorth}, J., {M{\o}ller}, P., {Gorosabel}, J., {et~al.} 2003{\natexlab{a}},
  \apj, 597, 699

\bibitem[{{Hjorth} {et~al.}(2003{\natexlab{b}}){Hjorth}, Sollerman, M{\o}ller,
  Fynbo, Woosley, Kouveliotou, Tanvir, Greiner, Andersen, Castro-Tirado, Cerón,
  Fruchter, Gorosabel, Jakbsson, Kaper, Klose, Masetti, Pedersen, Pedersen,
  Pian, Palazzi, Rhoads, Rol, van~den Heuvel, Vreeswijk, Watson, \&
  Wijers}]{hjorth03}
{Hjorth}, J., Sollerman, J., M{\o}ller, P., {et~al.} 2003{\natexlab{b}},
  Nature, 423, 847

\bibitem[{{Hjorth} {et~al.}(2002){Hjorth}, {Thomsen}, {Nielsen}, {Andersen},
  {Holland}, {Fynbo}, {Pedersen}, {Jaunsen}, {Halpern}, {Fesen}, {Gorosabel},
  {Castro-Tirado}, {McMahon}, {Hoenig}, {Bj{\" o}rnsson}, {Amati}, {Tanvir}, \&
  {Natarajan}}]{hjorth02}
{Hjorth}, J., {Thomsen}, B., {Nielsen}, S.~R., {et~al.} 2002, ApJ, 576, 113

\bibitem[{{Holland} {et~al.}(2001){Holland}, {Fynbo}, {Hjorth}, {Gorosabel},
  {Pedersen}, {Andersen}, {Dar}, {Thomsen}, {M{\o}ller}, {Bj{\" o}rnsson},
  {Jaunsen}, {Natarajan}, \& {Tanvir}}]{holland01}
{Holland}, S., {Fynbo}, J.~P.~U., {Hjorth}, J., {et~al.} 2001, A\&A, 371, 52

\bibitem[{{Holland} \& {Hjorth}(1999)}]{holland99b}
{Holland}, S. \& {Hjorth}, J. 1999, A\&A, 344, L67

\bibitem[{{Jakobsson} {et~al.}(2003){Jakobsson}, {Hjorth}, {Fynbo},
  {Gorosabel}, {Pedersen}, {Burud}, {Levan}, {Kouveliotou}, {Tanvir},
  {Fruchter}, {Rhoads}, {Grav}, {Hansen}, {Michelsen}, {Andersen}, {Jensen},
  {Pedersen}, {Thomsen}, {Weidinger}, {Bhargavi}, {Cowsik}, \&
  {Pandey}}]{jakobsson03}
{Jakobsson}, P., {Hjorth}, J., {Fynbo}, J.~P.~U., {et~al.} 2003, \aap, 408, 941

\bibitem[{{Jaunsen} {et~al.}(2003){Jaunsen}, {Andersen}, {Hjorth}, {Fynbo},
  {Holland}, {Thomsen}, {Gorosabel}, {Schaefer}, {Bjornsson}, {Natarajan}, \&
  {Tanvir}}]{jaunsen03}
{Jaunsen}, A.~O., {Andersen}, M.~I., {Hjorth}, J., {et~al.} 2003, A\&A, 402,
  125

\bibitem[{{Kennicutt}(1998)}]{kennicutt98}
{Kennicutt}, R.~C. 1998, ARA\&A, 36, 189

\bibitem[{{Kulkarni} {et~al.}(1998){Kulkarni}, {Frail}, {Wieringa}, {Ekers},
  {Sadler}, {Wark}, {Higdon}, {Phinney}, \& {Bloom}}]{kul98b}
{Kulkarni}, S.~R., {Frail}, D.~A., {Wieringa}, M.~H., {et~al.} 1998, Nature,
  395, 663

\bibitem[{{Le Floc'h} {et~al.}(2003){Le Floc'h}, {Duc}, {Mirabel}, {Sanders},
  {Bosch}, {Diaz}, {Donzelli}, {Rodrigues}, {Courvoisier}, {Greiner},
  {Mereghetti}, {Melnick}, {Maza}, \& {Minniti}}]{lefloch03}
{Le Floc'h}, E., {Duc}, P.-A., {Mirabel}, I.~F., {et~al.} 2003, \aap, 400, 499

\bibitem[{{Lilly} {et~al.}(1995){Lilly}, {Tresse}, {Hammer}, {Crampton}, \& {Le
  Fevre}}]{lilly95}
{Lilly}, S.~J., {Tresse}, L., {Hammer}, F., {Crampton}, D., \& {Le Fevre}, O.
  1995, ApJ, 455, 108

\bibitem[{{Madau} {et~al.}(1998){Madau}, {Pozzetti}, \& {Dickinson}}]{madau98}
{Madau}, P., {Pozzetti}, L., \& {Dickinson}, M. 1998, ApJ, 498, 106

\bibitem[{{Malesani} {et~al.}(2004){Malesani}, {Tagliaferri}, {Chincarini},
  {Covino}, {Della Valle}, {Fugazza}, {Mazzali}, {Zerbi}, {D'Avanzo},
  {Kalogerakos}, {Simoncelli}, {Antonelli}, {Burderi}, {Campana}, {Cucchiara},
  {Fiore}, {Ghirlanda}, {Goldoni}, {G{\" o}tz}, {Mereghetti}, {Mirabel},
  {Romano}, {Stella}, {Minezaki}, {Yoshii}, \& {Nomoto}}]{malesani04}
{Malesani}, D., {Tagliaferri}, G., {Chincarini}, G., {et~al.} 2004, \apjl, 609,
  L5

\bibitem[{{Miller} \& {Scalo}(1979)}]{miller79}
{Miller}, G.~E. \& {Scalo}, J.~M. 1979, ApJS, 41, 513

\bibitem[{{Park} {et~al.}(2002){Park}, {Williams}, {Hartmann}, {Lamb}, {Lee},
  {Tucker}, {Klose}, {Stecklum}, {Henden}, {Adelman}, {Barthelmy}, {Briggs},
  {Brinkmann}, {Chen}, {Cline}, {Csabai}, {Gehrels}, {Harvanek}, {Hennessy},
  {Hurley}, {Ivezi{\' c}}, {Kent}, {Kleinman}, {Krzesinski}, {Lindsay}, {Long},
  {Nemiroff}, {Neilsen}, {Nitta}, {Newberg}, {Newman}, {Perez}, {Periera},
  {Schneider}, {Snedden}, {Stoughton}, {Vanden Berk}, {York}, \&
  {Ziock}}]{park02}
{Park}, H.~S., {Williams}, G.~G., {Hartmann}, D.~H., {et~al.} 2002, \apjl, 571,
  L131

\bibitem[{{Peacock}(1983)}]{peacock83}
{Peacock}, J.~A. 1983, \mnras, 202, 615

\bibitem[{{Piro} {et~al.}(2002){Piro}, {Frail}, {Gorosabel}, {Garmire},
  {Soffitta}, {Amati}, {Andersen}, {Antonelli}, {Berger}, {Frontera}, {Fynbo},
  {Gandolfi}, {Garcia}, {Hjorth}, {Zand}, {Jensen}, {Masetti}, {M{\o}ller},
  {Pedersen}, {Pian}, \& {Wieringa}}]{piro02}
{Piro}, L., {Frail}, D.~A., {Gorosabel}, J., {et~al.} 2002, \apj, 577, 680

\bibitem[{{Press} {et~al.}(1992){Press}, {Teukolsky}, {Vetterling}, \&
  {Flannery}}]{num}
{Press}, W.~H., {Teukolsky}, S.~A., {Vetterling}, W.~T., \& {Flannery}, B.~P.
  1992, {Numerical recipes in FORTRAN. The art of scientific computing}
  (Cambridge: University Press, 1992, 2nd ed.)

\bibitem[{{Prevot} {et~al.}(1984){Prevot}, {Lequeux}, {Prevot}, {Maurice}, \&
  {Rocca-Volmerange}}]{prevot84}
{Prevot}, M.~L., {Lequeux}, J., {Prevot}, L., {Maurice}, E., \&
  {Rocca-Volmerange}, B. 1984, A\&A, 132, 389

\bibitem[{{Price} {et~al.}(2002){Price}, {Kulkarni}, {Berger}, {Djorgovski},
  {Frail}, {Mahabal}, {Fox}, {Harrison}, {Bloom}, {Yost}, {Reichart}, {Henden},
  {Ricker}, {van der Spek}, {Hurley}, {Atteia}, {Kawai}, {Fenimore}, \&
  {Graziani}}]{price02}
{Price}, P.~A., {Kulkarni}, S.~R., {Berger}, E., {et~al.} 2002, \apjl, 571,
  L121

\bibitem[{{Prochaska} {et~al.}(2004){Prochaska}, {Bloom}, {Chen}, C.,
  {Melbourne}, {Dressler}, {Graham}, \& {Osip}}]{prochaska04}
{Prochaska}, J.~X., {Bloom}, J.~S., {Chen}, H., {et~al.} 2004, ApJ in press.

\bibitem[{{Salpeter}(1955)}]{salpeter55}
{Salpeter}, E.~E. 1955, \apj, 121, 161

\bibitem[{{Schechter}(1976)}]{schechter}
{Schechter}, P. 1976, ApJ, 203, 297

\bibitem[{{Schlegel} {et~al.}(1998){Schlegel}, {Finkbeiner}, \&
  {Davis}}]{schlegel98}
{Schlegel}, D.~J., {Finkbeiner}, D.~P., \& {Davis}, M. 1998, ApJ, 500, 525

\bibitem[{{Seaton}(1979)}]{seaton79}
{Seaton}, M.~J. 1979, MNRAS, 187, 73P

\bibitem[{{Sokolov} {et~al.}(2001){Sokolov}, {Fatkhullin}, {Castro-Tirado},
  {Fruchter}, {Komarova}, {Kasimova}, {Dodonov}, {Afanasiev}, \&
  {Moiseev}}]{sok01}
{Sokolov}, V.~V., {Fatkhullin}, T.~A., {Castro-Tirado}, A.~J., {et~al.} 2001,
  A\&A, 372, 438

\bibitem[{{Stanek} {et~al.}(2003){Stanek}, {Matheson}, {Garnavich}, {Martini},
  {Berlind}, {Caldwell}, {Challis}, {Brown}, {Schild}, {Krisciunas}, {Calkins},
  {Lee}, {Hathi}, {Jansen}, {Windhorst}, {Echevarria}, {Eisenstein}, {Pindor},
  {Olszewski}, {Harding}, {Holland}, \& {Bersier}}]{stanek03}
{Stanek}, K.~Z., {Matheson}, T., {Garnavich}, P.~M., {et~al.} 2003, \apjl, 591,
  L17

\bibitem[{{Tanvir} {et~al.}(2004){Tanvir}, {Barnard}, {Blain}, {Fruchter},
  {Kouveliotou}, {Natarajan}, {Ramirez-Ruiz}, {Rol}, {Smith}, {Tilanus}, \&
  {Wijers}}]{tanvir04}
{Tanvir}, N.~R., {Barnard}, V.~E., {Blain}, A.~W., {et~al.} 2004, MNRAS, in
  press.

\bibitem[{{Vanzella} {et~al.}(2001){Vanzella}, {Cristiani}, {Saracco},
  {Arnouts}, {Bianchi}, {D'Odorico}, {Fontana}, {Giallongo}, \&
  {Grazian}}]{vanzella01}
{Vanzella}, E., {Cristiani}, S., {Saracco}, P., {et~al.} 2001, \aj, 122, 2190

\bibitem[{{Vreeswijk} {et~al.}(2001){Vreeswijk}, {Fender}, {Garrett}, {Tingay},
  {Fruchter}, \& {Kaper}}]{vrees01b}
{Vreeswijk}, P.~M., {Fender}, R.~P., {Garrett}, M.~A., {et~al.} 2001, A\&A,
  380, L21

\bibitem[{{Vreeswijk} {et~al.}(1999){Vreeswijk}, {Galama}, {Owens},
  {Oosterbroek}, {Geballe}, {van Paradijs}, {Groot}, {Kouveliotou}, {Koshut},
  {Tanvir}, {Wijers}, {Pian}, {Palazzi}, {Frontera}, {Masetti}, {Robinson},
  {Briggs}, {in 't Zand}, {Heise}, {Piro}, {Costa}, {Feroci}, {Antonelli},
  {Hurley}, {Greiner}, {Smith}, {Levine}, {Lipkin}, {Leibowitz}, {Lidman},
  {Pizzella}, {B{\" o}hnhardt}, {Doublier}, {Chaty}, {Smail}, {Blain}, {Hough},
  {Young}, \& {Suntzeff}}]{vrees99}
{Vreeswijk}, P.~M., {Galama}, T.~J., {Owens}, A., {et~al.} 1999, ApJ, 523, 171

\bibitem[{{Woosley}(1993)}]{woosley93}
{Woosley}, S.~E. 1993, \apj, 405, 273

\end{thebibliography}
\end{document}